\documentstyle[11pt,aaspp4]{article}

\received{}
\journalid{}{}
\articleid{}{}

\slugcomment{}

\begin{document}

\title{THE {\it IRAS} 1-JY SAMPLE OF ULTRALUMINOUS INFRARED GALAXIES: \break
II. OPTICAL SPECTROSCOPY\footnote{This work was part of a Ph.D. thesis by 
D.--C. Kim completed in the Department of Physics and Astronomy, 
University of Hawaii, Honolulu, HI}}

\author{D.-C. Kim,\altaffilmark{2,3} Sylvain Veilleux,\altaffilmark{4}
and D. B. Sanders\altaffilmark{5}}

\altaffiltext{2}{Current address: Infrared Processing and Analysis Center, 
California Institute of Technology, Pasadena, CA  91125;  
E-mail: kim@ipac.caltech.edu}
\altaffiltext{3}{Institute for Astronomy, University of Hawaii, 2680 Woodlawn Drive,
Honolulu, HI 96822}
\altaffiltext{4}{Department of Astronomy, University of Maryland, 
College Park, MD 20742; E-mail: veilleux@astro.umd.edu}
\altaffiltext{5}{Institute for Astronomy, University of Hawaii, 2680 Woodlawn Drive,
Honolulu, HI 96822; sanders@ifa.hawaii.edu}

\newcommand{\ts}{\thinspace}        

\begin{abstract}
This is the second paper in a series discussing the properties of
ultraluminous infrared galaxies (ULIGs: {\it L}$_{\rm ir}$ $>$
10$^{12}$ {\it L}$_\odot$; {\it H}$_o$ = 75 km s$^{-1}$ Mpc$^{-1}$ and
{\it q}$_o$ = 0.0) from the 1-Jy sample of Kim (1995). This paper
presents the first results of a spectroscopic survey at optical
wavelengths of a randomly selected subset of 45 ULIGs from Kim \& Sanders 
(1997).  These new data are combined with previous data from Veilleux et
al. (1995) to determine the spectral properties of luminous infrared
galaxies (LIGs) with $L_{\rm ir} \approx 10^{10.5}-10^{13}\ L_\odot$.  
We find that the fraction
of Seyfert galaxies among LIGs increases
dramatically above $L_{\rm ir} \approx 10^{12.3}\ L_\odot$ --- 
nearly half of the galaxies
with $L_{\rm ir} > 10^{12.3}\ L_\odot$ present Seyfert characteristics. 
Many of the
optical properties of these Seyfert galaxies are consistent with the
presence of a genuine active galactic nucleus (AGN) in the core of
these objects.

The continuum colors and strengths of the stellar H$\beta$ and Mg{\ts}Ib
features in and out of the nuclei of ULIGs indicate that 
star formation has recently ($\sim${\ts}10$^7$ yr) taken place in the
nuclear and circumnuclear regions of many of these objects. 
As expected, 
photoionization by hot stars appears to be the dominant source of ionization
in the objects with H{\ts}II region-like spectra.
Evidence is presented that the ionization source in
infrared-selected galaxies with nuclear LINER-like spectra (38\% 
of the ULIGs in our sample) is likely to be shocks or of stellar
origins rather than an AGN. Shock ionization associated with 
starburst-driven outflows may also explain the LINER-like
emission detected {\it outside} the nuclei of some galaxies. 

No significant differences are found between the mean color excess of
ULIGs and that of {\it IRAS} galaxies of lower infrared
luminosity. However, in constrast to what was found in low-luminosity
infrared galaxies, the color excess in the nuclei of ULIGs does not
seem to depend on spectral types.  The reddening in ULIGs is generally
observed to decrease with distance from the nucleus as in their
low-luminosity counterparts. Clear cases of inverted reddening
profiles are observed in less than 1/5 of the ULIGs in our sample,
nearly all of which are optically classified LINERs.  These profiles may
reflect the actual dust distribution in these objects or may be due to
complex optical depth effects.
\end{abstract}


\section{Introduction}
Over the last decade, several spectroscopic studies have attempted to determine 
the nature of the dominant energy source in ultraluminous infrared galaxies 
(ULIGs: $L_{\rm ir} > 10^{12}\ L_\odot$)\footnote{$L_{\rm ir} \equiv L (8-1000 \micron)$, 
computed from the observed infrared fluxes in all four {\it IRAS} bands according to 
the prescription outlined in Perault (1987).  We assume $H_{\rm o} = 75$ km{\ts}s$^{-1}${\ts}Mpc$^{-1}$ and $q_{\rm o} = 0$ throughout this paper.}.  
Spectroscopic surveys of unbiased samples of luminous {\it IRAS} galaxies 
(e.g. Elston, Cornell, \& Lebofsky 1985; Leech et al. 1989; Allen et al. 1991;  
Ashby, Houck, \& Matthews 1995) generally have found that $\gtrsim${\ts}80\% 
of high-luminosity infrared galaxies ($L_{\rm ir} > 10^{11}\ L_\odot$) 
present H{\ts}II region-like optical spectra, and therefore appear to be powered 
by hot stars rather than an active galactic nucleus (AGN).  However, the 
great majority of the infrared galaxies in these
samples are in the luminosity range $L_{\rm ir} = 10^{11}-10^{12}\ L_\odot$, 
with only a few having $L_{\rm ir} > 10^{12}\ L_\odot$. This distinction is important 
as there is growing evidence that the fraction of AGN among LIGs increases with
increasing $L_{\rm ir}$ (see review by Sanders \& Mirabel 1996).  An early 
analysis by Sanders et al. (1988) of 10 ULIGs in the Bright Galaxy Survey 
(BGS: Soifer et al. 1987, 1989) suggested that 9 of these objects had 
AGN-like optical spectra (i.e. either Seyferts or LINERs).
In addition, Armus, Heckman \& Miley (1989) found that 10 of the 12 
ULIGs in their sample of slightly more distant objects presented AGN-like spectra.  
More recently, Veilleux et al. (1995; hereafter VKSMS) carried out a 
sensitive spectroscopic survey of a sample of 200
LIGs and classified the nuclear spectra of these
galaxies using a large number of line-ratio diagnostics corrected for
the underlying stellar absorption features.  They found that 13 of the
21 ULIGs in their sample (including the 10 ULIGs from the BGS) showed
AGN-like spectra.

Although the results from previous spectroscopic surveys of ULIGs are 
important, the relatively small number of objects studied, the limited range 
of $L_{\rm ir}$ of these objects (nearly all have $L_{\rm ir} < 10^{12.3}\ L_\odot$), and the 
relatively low resolution and modest signal-to-noise of many of the spectra,  
make it desirable to obtain more sensitive and higher resolution data 
for a larger sample of objects.  The recent publication of the 1-Jy 
sample of ULIGs (Kim \& Sanders 1997; hereafter Paper I) provides such a 
sample that can now be used for a more complete study of the 
properties of ULIGs.  The 1-Jy survey contains a complete 
sample of 111 ULIGs ($\vert b \vert > 30^\circ$, $\delta > -40^\circ$) plus a 
smaller supplementary list of 8 ULIGs at lower Galactic latitude 
($\vert b \vert = 20^\circ - 30^\circ$) that were selected from the {\it IRAS} 
Faint Source Catalog--Version 2 (FSC: Moshir et al. 1992) on the basis of their 60\micron\ flux: 
${\it f}_{60} > 1$-Jy\footnote{The quantities ${\it f}_{12}$, ${\it f}_{25}$, 
${\it f}_{60}$, ${\it f}_{100}$, are the {\it IRAS} flux densities in Jy at 
12\micron, 25\micron, 60\micron, and 100\micron\ respectively} (see Paper I for a 
complete description of the sample and selection criteria). 

This paper presents the first results from a spectral analysis of 
a randomly selected subset of 45 ULIGs selected from Paper I.  
Spectra on the remaining objects in the 1-Jy sample have been collected
and are presently being analyzed (Veilleux, Kim, \& Sanders 1997a). 
The current subsample, which includes 35{\ts}\% of the total number of 
objects in the complete 1-Jy survey, was selected to be fairly evenly distributed
in redshift, and should be representative of the complete 1-Jy sample.
Unlike previous spectroscopic studies of ULIGs, where the majority of
objects have $L_{\rm ir} < 10^{12.3}\ L_\odot$, the current survey has
a large fraction of objects ($\sim${\ts}45{\ts}\%) with higher luminosity ($L_{\rm
ir} = 10^{12.3}-10^{12.9}\ L_\odot$), and thus provides for the first
time a more realistic picture of the spectroscopic properties of ULIGs
throughout the decade luminosity range $L_{\rm ir} = 10^{12}-10^{13}\
L_\odot$.  Section 2 discusses the procedures used to obtain and
reduce the new data.  Details of the line and continuum measurements
are discussed in \S 3.  The spectral analysis was carried out in both
the nuclear and circumnuclear regions of these galaxies.  The results
derived from the nuclear spectra are described in \S 4, while the
radial profiles of a number of spectral parameters are analyzed in \S
5. The main conclusions are summarized in \S 6.

\section{Observations and Data Reduction}

All of the spectroscopic data were obtained with the Faint Object
Spectrograph at the f/10 Cassegrain focus of the University of Hawaii
2.2-m telescope on Mauna Kea.  Table 1 lists the dates of the
observations, grating used, spectral coverage, resolution, seeing, and
detectors used for the observations.  The exposure times range from
300 sec to 2,000 sec depending on the $R$ magnitude of each object
(see Kim, Sanders \& Mazzarella 1996; hearafter Paper III).  In all
cases, a slit with a width of 2\arcsec\ was used and the slit was
positioned in the E-W direction ($PA = 90^\circ$).  All observations
were made under photometric conditions.

The reduction techniques of Kim et al. (1995) were used to reduce the
present data.  To minimize aperture-related effects, the window used
for the extraction of the nuclear spectrum was varied according to the
redshift of each object so that it corresponds to a constant linear
scale (total diameter) of 4 kpc for most of the galaxies, with the
exceptions of five distant galaxies with $z$ $>$ 0.15 (FSC{\ts}10091+4704,
FSC{\ts}11582+3020, FSC{\ts}13218+0552, FSC{\ts}14070+0525, and FSC{\ts}16300+1558)\footnote{Object 
names that begin with `FSC' are sources identified in the {\it IRAS} Faint 
Source Catalog--Version 2 (Moshir et al. 1992)} for which an
8-kpc window was used. Note, however, that the effective size of the
aperture still has a dependence on redshift since the slit width was
held constant at 2\arcsec\ during all of the observations.  The resulting
spectra contained absorption bands near 6870{\ts}\AA~and 7620{\ts}\AA~produced
by atmospheric O$_2$ (the B- and A-bands, respectively). These features
were removed using the spectra of the flux standard stars.

\section{Line and Continuum Measurements}
 
The final calibrated spectra are plotted in Figure 1.  Table 2 lists
the measured redshifts, the infrared luminosities computed from these
redshifts using the procedure outlined in Kim et al. (1995), and the
spectral type as determined from the spectra in Figure 1 (see \S 4.1).
Also listed for completeness in Table 2 are the basic properties of 7
other ULIGs from the 1-Jy sample for which spectroscopic data already
exist in the literature. However, the majority of these previously
studied objects are biased toward AGN---two objects were previously
identified in optical QSO surveys and 3 were selected on the basis of
``warm'' infrared colors, $f_{25}/f_{60} > 0.25$, which is known to
preselect objects with AGN-like optical spectra (e.g. deGrijp et al. 
1985).  Therefore, to avoid biasing our 1-Jy sample classifications
toward AGN, only the two BGS ULIGs out of
the 7 previously classified objects (these two
ULIGs are flagged with a dagger in Table 2) will be used
in this paper for computing the distribution of spectral types (\S
4.1), and since new spectra were not obtained for any of the 7
previously classified ULIGs none of these 7 objects will be included
in any other analyses.

Tables 3 and 4 list the various spectral parameters
measured from our new data.  Following Kim et al. (1995), the line
fluxes were measured using two methods. The fluxes of isolated
emission lines were measured using the standard plotting package in
IRAF (``splot''), while ``specfit'', an interactive spectral analysis
procedure linked to IRAF and kindly provided by Gerard A. Kriss, was
used to deal with blended lines (e.g., H$\alpha$ + [N{\ts}II]
$\lambda\lambda$6548, 6583 and [S{\ts}II] $\lambda\lambda$6716, 6731) and
emission lines affected by stellar absorption features (e.g. H$\beta$ and
H$\alpha$). This last routine can fit a wide variety of emission-line,
absorption-line, and continuum models to the observed spectrum.  The
input parameters for the fit were determined through ``splot'' in
IRAF.  We chose to fit the continuum with a simple first-order
polynomial, and the emission and absorption lines with Gaussian
profiles.  The actual fitting was done via a chi-square minimization
using a simplex algorithm.  The output parameters were the flux level
and slope of the underlying continuum emission, the flux, centroid,
and full-width-half-maximum (FWHM) of the emission lines, and equivalent 
width (EW), centroid, and {FWHM of the absorption lines. All of the 
EWs discussed in this paper were derived in the object rest frame. 

\section{Results from the Analysis of the Nuclear Data}

\subsection{Spectral Classification}

All of the ULIGs in our current study  
present a spectrum with emission lines.
The characteristics of these lines are important indicators of
the physical conditions of the thermal gas in our sample galaxies.
In an attempt to determine the dominant source of ionization in these
objects, four diagnostic line ratios 
known to be efficient at differentiating 
between the various ionization mechanisms (Veilleux \& Osterbrock 1987;
hereafter VO87) were used.
These are [O{\ts}III] $\lambda$5007/H$\beta$, [N{\ts}II] $\lambda$6583/H$\alpha$,
[S{\ts}II] $\lambda\lambda$6716,~6731/H$\alpha$, and [O{\ts}I] $\lambda$6300/H$\alpha$.
All of the line ratios used in the present analysis have been
corrected for reddening using the values of {\it E(B--V)} determined from
the H$\alpha$/H$\beta$ ratio and the
Whitford reddening curve as parameterized by Miller \& Mathews (1972;
see next section). 

The measurements are listed in Table 5 and are plotted in Figure 2.
For each object in Table 5, entries on the first row represent the
observed line ratios, while the second row lists the dereddened line
ratios.  Columns (8), (9), and (10) list the spectral types determined
from the location of the data points in the diagrams of [O{\ts}III]
$\lambda$5007/H$\beta$ versus [N{\ts}II] $\lambda$6583/H$\alpha$, [S{\ts}II]
$\lambda\lambda$6716, 6731/H$\alpha$, and [O{\ts}I]
$\lambda$6300/H$\alpha$, respectively.  The boundaries of VO87 were used 
to classify each object as H{\ts}II or AGN-like galaxies.  A distinction
was made among AGN-like galaxies between the objects of high ([O{\ts}III]
$\lambda$5007/H$\beta \ge$ 3) and low ([O{\ts}III] $\lambda$5007/H$\beta
\leq$ 3) excitation.  The first group represents the ``classic''
Seyfert 2 galaxies while galaxies in the second group were classified
as LINERs (``low-ionization nuclear emitting regions''), although a
few of them may not satisfy the original LINER definition of Heckman
(1980).  Finally, column (11) of Table 5 gives the adopted spectral
type based on the previous three columns.  Galaxies with Fe II
multiplets at $\lambda\lambda$5100--5560 and very broad 
($\Delta V_{\rm FWHM} \gtrsim 1000$ km s$^{-1}$) H{\ts}I Balmer and 
He{\ts}I $\lambda$5876 emission
lines were classified as Seyfert 1s. Eight systems in our sample
appear double. In FSC{\ts}08572+3915 (LINER, LINER), FSC{\ts}14348--1447 (LINER,
LINER), FSC{\ts}16468+5200 (LINER, LINER), and FSC{\ts}17028+5817 (LINER, H{\ts}II), 
a LINER classification was adopted for the whole system, while in
FSC{\ts}14202+2615 (H{\ts}II, main nucleus), FSC{\ts}14394+5332 (Seyfert 2, western
nucleus), FSC{\ts}16333+4630 (LINER, western main nucleus), and FSC{\ts}18470+3233
(H{\ts}II, main component on the south-west), only the spectrum of
the main component was available, and therefore the spectral type of that
component was adopted for these systems.

Out of the 47 ULIGs that make up our unbiased infrared selected sample
(i.e. all sources in Table 2 except the 5 sources identified in previous 
optically selected AGN catalogs), 28\% of them (13/47) were found to have spectra
characteristic of photoionization by hot stars (H{\ts}II region-like).
AGN-like emission lines were observed in 72\% (34/47) of the
total sample including 13\% (6/47) Seyfert 1s, 21\% (10/47) Seyfert 2s, 
and 38\% (18/47) LINER-like objects.  These results
were combined with the measurements obtained by VKSMS for the LIGs
 from the BGS to search for systematic variations of
the spectral types with infrared luminosity. For this exercise, ULIGs
were further divided into two luminosity bins: $10^{12}\ L_\odot \leq L_{\rm ir}
 < 10^{12.3}\ L_\odot$ and $L_{\rm ir} \geq 10^{12.3}\ L_\odot$.  
A summary of this analysis is given in
Table 6 and plotted in Figure 3. There is an obvious tendency for the
more luminous objects to have AGN-like line ratios and for these
objects to be more Seyfert-like.  In fact, none of the Seyfert 1
galaxies have $L_{\rm ir} < 10^{11}\ L_\odot$ 
and about half of the galaxies with
$L_{\rm ir} \geq 10^{12.3}\ L_\odot$ present Seyfert characteristics.

LINER-like galaxies constitute 30--40\% of the total
sample regardless of $L_{\rm ir}$.  Essentialy all of the
ULIGs in our sample classified as H{\ts}II galaxies have relatively low
ionization level ([O{\ts}III] $\lambda$5007/H$\beta$ $<$ 3) and fall
outside of the region populated by galaxies with very recent bursts of
star formation (Evans \& Dopita 1985; McCall, Rybski, \& Shields 1985;
Allen et al. 1991).  The only galaxy that lies anywhere near this
region of the [O{\ts}III]/H$\beta$ vs. [N{\ts}II]/H$\alpha$ diagram is
FSC{\ts}15206+3342.  This general lack of high-excitation H{\ts}II galaxies among
LIGs was also observed in the samples of VKSMS,
Leech et al. (1989), Armus et al. (1989), Allen et al. (1991), Ashby, Houck, \&
Hacking (1992), and Ashby et al. (1995).  The lack of Wolf-Rayet
features in our sample galaxies also argues against a very recent
burst ($<<$ 10$^7$ yr) of star formation.

As indicated in the reference column of Table 2, seven galaxies 
from the sample of VKSMS have also
been observed in the present study. The spectral types derived from these
two sets of data agree for all but two of these objects: FSC{\ts}13428+5606 (now 
classified as a Seyfert 2 galaxy rather than a LINER) and 
FSC{\ts}15327+2340 (now classified as a LINER instead of a Seyfert 2 galaxy). 
The different classification of the first object is due to
a slight change in the [O{\ts}III]~$\lambda$5007/H$\beta$ ratio which may be 
attributable to differences in the sizes of the extraction apertures used
for these data. The present data were extracted from the central 4 kpc 
diameter region of this
galaxy while  VKSMS used a 2 kpc aperture. 
The difference in the spectral type of FSC{\ts}15327+2340 is probably
due to the rather poor quality of the original  VKSMS data on this object 
compounded with centering errors more likely to affect the older data. 
For both of these galaxies, the spectral type derived from the 
present data will be used in the following discussion.

\subsection{Reddening}

The reddening in our sample galaxies was estimated using the
emission-line H$\alpha$/H$\beta$ ratios
corrected for the underlying stellar absorption features. 
An intrinsic H$\alpha$/H$\beta$
ratio of 2.85 was adopted for H{\ts}II region-like galaxies (Case B Balmer recombination
decrement for $T = 10^4$ K and $n_{\rm e} = 10^4$ cm$^{-3}$) and 3.10 for
AGN-like galaxies (see references in  VO87).
The Whitford reddening curve as parameterized by Miller \&
Mathews (1972) was used. 
The reddenings derived from this method are listed as color
excesses, {\it E(B--V)}, in the third column of Table 5. 
Since nearly all the objects in our sample lie at $|b| > 30 ^\circ$,
no correction was made for Galactic reddening.
Note that the extinction measurements based on the
optical spectra are likely to represent lower limits to the true extinction
because the observed line and continuum emission from an inhomogeneous
environment like that of the nuclear regions of ULIGs will come
predominantly from regions of smaller optical depths.

The distribution of {\it E(B--V)} for the ULIGs of our sample is shown
in Figure 4.  The median {\it E(B--V)} are 1.13, 1.02 and 1.54 for
H{\ts}II galaxies, LINERs and Seyfert 2 galaxies, respectively.
Kolmogorov-Smirnov (K-S) tests indicate that the differences between
the various spectral types are not significant. These color excesses
are similar to those obtained by VKSMS in the lower-luminosity
galaxies of the BGS [{\it E(B--V)} = 1.05, 1.24, and 1.07 for H{\ts}II
galaxies, LINERs, and Seyfert 2 galaxies, respectively]. However,
VKSMS found that the color excesses of the LINERs were significantly
larger than those of the H{\ts}II and Seyfert 2 galaxies.  The color
excesses measured in the ULIGs of our sample and the lower luminosity
objects of VKSMS are considerably larger than those measured in
optically-selected Seyfert and starburst galaxies (Dahari \&
De~Robertis 1988) and in extragalactic H{\ts}II regions (Kennicutt, Keel,
\& Blaha 1989).  These results confirm the importance of dust in
infrared galaxies, regardless of their infrared luminosity.

The correlation observed by VKSMS between {\it E(B--V)} and {\it
EW}(Na{\ts}Id) of H{\ts}II galaxies is also present, but perhaps at a weaker
level, among our high-luminosity objects (Fig. 5).  The probabilities,
P[null], that these correlations are fortuitous are $3\times10^{-4}$,
0.07, and 0.10 among the H{\ts}II galaxies, LINERs, and Seyfert 2 galaxies
of the combined sample.  Figure 6 shows that a similar correlation is
observed between {\it E(B--V)} and the observed continuum colors of H{\ts}II
galaxies ($P{\rm [null]} = 2 \times 10^{-11}$) and, to a lesser extent,
LINERs ($P{\rm [null]} = 2 \times 10^{-4}$), but not among the Seyfert 2
galaxies ($P{\rm [null]} = 0.78$).  As first reported by VKSMS, these results
suggest that the scatter in the Na{\ts}Id equivalent widths and continuum
colors of the Seyfert LIGs is predominantly intrinsic rather than
caused by variations in the amount of reddening from one object to the
other.

\subsection{Densities}

Figure 7 shows the distribution of the electron densities 
as a function of spectral type.
K-S tests indicate that these distributions are not significantly different.
The median values of the electron densities, $n_{\rm e}$, 
are 170 cm$^{-3}$, 140 cm$^{-3}$, and 
280 cm$^{-3}$ for H{\ts}II region galaxies, LINERs and Seyfert 2 galaxies,
respectively. These median densities
are comparable to those obtained by VKSMS for the lower luminosity
infrared galaxies, but are considerably smaller than those found 
by Koski (1978) and Keel (1983) 
in optically-selected LINERs, Seyfert 2 and
narrow-line radio galaxies. 
A possible explanation for this difference is that the
strong dust extinction in infrared-selected galaxies 
prevents us from probing the inner, presumably denser 
[S{\ts}II] line-emitting zone in these objects (see VKSMS). 

\subsection{Line Widths}

[O{\ts}III] $\lambda$5007 was selected for line width
measurements because it is strong in most ULIGs and free of any nearby 
emission or absorption lines (in contrast to H$\alpha$). The line
widths listed in Table 4 have been corrected for the
finite instrument resolution of the data ($\sim${\ts}8{\ts}\AA) using the
quadrature method. This method assumes
that the intrinsic and instrument profiles are Gaussian and gives
corrected widths that are too large for profiles which
are more peaky than Gaussians (e.g., the emission-line profiles in AGN;
Whittle 1985; Veilleux 1991). For this
reason, the [O{\ts}III] $\lambda$5007 line widths should be treated with
caution. 

Figure 8 shows the distribution of the 
[O{\ts}III] line widths for each spectral type.
The median line widths, $\Delta V_{\rm FWHM}$, of the H{\ts}II galaxies and LINERs are 
comparable (170 km s$^{-1}$ and 130 km s$^{-1}$) whereas the [O{\ts}III] line 
widths of Seyfert 2 galaxies appear larger (500 km s$^{-1}$). K-S
tests indicate, however, that the differences are not particularly significant
because of the small number of galaxies in each category.
The probability that the line widths
of H{\ts}II galaxies and LINERs are drawn from the same population is
0.40, it is 0.04 when comparing the line widths of 
H{\ts}II and Seyfert 2 galaxies, and 0.10 for the 
LINERs and Seyfert 2 galaxies.
In comparison, VKSMS derived median line widths of 255, 310, and 350 km
s$^{-1}$ for the H{\ts}II, LINER, and Seyfert 2 LIGs of the BGS. A K-S comparison
of these results with those that were obtained for our ULIGs
show that they are not statistically very different.

Based on the existence of objects with very
broad and complex line profiles in their sample, VKSMS 
argued that the line 
broadening in some LIGs (particularly those with AGN-like spectral features)
is caused in part by non-gravitational processes associated with
nuclear outflows. Figure 9 illustrates the distribution of [O{\ts}III]
line widths of the combined sample as a function of infrared
luminosity. It is interesting to note that all the objects with line widths larger than
600 km s$^{-1}$ have $L_{\rm ir} \gtrsim 10^{11}\ L_\odot$, 
and objects with the most extreme
profiles ($\Delta V_{\rm FWHM} > 1000$ km s$^{-1}$) all have $L_{\rm ir} > 10^{ 12}\ L_\odot$ and Seyfert
characteristics. When the line widths measured in both samples are plotted 
as a function of the {\it IRAS} flux
density ratio ${\it f}_{25}/{\it f}_{60}$, a well-known indicator of Seyfert activity in
infrared galaxies (e.g., deGrijp et al. 1985; Miley, Neugebauer \& Soifer 1985),
a significant correlation is found among Seyfert 2 galaxies 
($P{\rm [null]} = 0.01$) but none among H{\ts}II galaxies ($P{\rm [null]} = 0.36$) or
LINERs ($P{\rm [null]} = 0.22$). These results (see Figure 10) suggest that the nuclear
activity in the Seyfert ULIGs may contribute to the line broadening. 

\subsection{Stellar Absorption Features}

The distributions of the equivalent widths of H$\beta$ and Mg{\ts}Ib are
presented in Figures 11 and 12. Unfortunately, our spectra do not
extend far enough in wavelengths to cover the important Ca{\ts}II triplet
feature (see VKSMS).  The median {\it EW}(H$\beta$) for our ULIGs are
0.77{\ts}\AA, 0.84{\ts}\AA, and 0.41{\ts}\AA\ for the H{\ts}II galaxies,
LINERs, and Seyfert 2 galaxies, respectively. The small number of
Seyfert galaxies with accurately measured H$\beta$ makes the
apparently smaller equivalent width of these objects uncertain. K-S
tests indicate, however, that these equivalent widths are
significantly smaller than those measured by VKSMS in the BGS LIGs  
(2.28{\ts}\AA, 2.18{\ts}\AA, and 2.49{\ts}\AA, respectively;
$P{\rm [null]} = 0.03$, 0.4, 0.005, respectively).  
The intermediate-age (10$^8$--10$^9$ yr) population of stars that is
responsible for the H$\beta$ absorption feature is less prominent in
our sample of ULIGs. A possibility is that the H$\beta$ feature is
being diluted by a hot, featureless continuum from young stars
(particularly in H{\ts}II galaxies) or from an AGN (in Seyfert 2
galaxies).

In this context, it is not surprising to find that
the {\it EW}(Mg Ib) for our sample of ULIGs
(median values of 1.1{\ts}\AA, 1.4{\ts}\AA, and 0.7{\ts}\AA~for H{\ts}II galaxies, 
LINERs, and Seyfert 2 galaxies, respectively)
are considerably smaller that those measured in 
non-active spiral galaxies [{\it EW}(Mg{\ts}Ib) = 3.5--5.0{\ts}\AA: Keel 1983; 
Stauffer 1982; Heckman, Balick, \& Crane 1980]. 
However, these values are {\it not} 
significantly different from those measured in 
the BGS objects of  VKSMS (1.12{\ts}\AA, 1.49{\ts}\AA, and 1.17{\ts}\AA,
respectively). Consequently,
the stellar population of LIGs {\it and} ULIGs is probably not
old enough ($>$ 10$^9$ yr; Bica, Alloin, \& Schmidt  1990) 
to produce strong {\it EW}(Mg{\ts}Ib). Any featureless continuum from young
stars or from an AGN would also weaken 
the spectral signatures of the underlying old stellar population. 
This issue will be discussed further when the radial behavior of these spectral
features in the sample galaxies is presented (\S 5.4). 

\subsection{Continuum Colors}

The observed continuum colors were dereddened 
using the amount of dust derived from the emission-line
spectrum. The results are presented in Figure 13. 
The dereddened continuum colors of the
Seyfert 2 ULIGs may be slightly bluer on average 
than those of the H{\ts}II galaxies and LINERs but
the difference is not statistically significant 
(median [C6563/C4861]$_0$ = 0.39, 0.39, and 0.22, for H{\ts}II galaxies, LINERs,
and Seyfert 2 galaxies, respectively).
In Figure 14 shows a strong anti-correlation between the 
dereddened continuum
colors and H$\alpha$ luminosities in the combined sample of
LIGs ($P{\rm [null]} = 2 \times 10^{-6}$, $9 \times 10^{-5}$, and 
0.005 for the H{\ts}II galaxies, LINERs, and Seyfert 2
galaxies, respectively). 
Interestingly, this anti-correlation virtually disappears ($P{\rm [null]} \gtrsim 0.01$) 
when the infrared luminosities are used instead of 
the H$\alpha$ luminosities (Figs. 15), 
suggesting that the driving parameter that determines
the optical continuum color in these galaxies 
is the luminosity that emerges from the
nucleus at optical wavelengths rather than the total energy output. 
The anti-correlation observed in Figure 14 may also be an artifact of our 
assumption that the continuum emission is reddened by the same amount as the 
emission lines.  Although it is unlikely that the amount of 
reddening in the emission-line gas has be overestimated (\S 4.2), the same may not be true of 
the continuum source.  Indeed, the inverse correlation shown in Figure 14 could 
result if the ionized gas is more strongly concentrated than the continuum 
source (in which case the dereddened continuum would be overcorrected for 
reddening and therefore too blue), and if the relative concentration of the 
ionized gas increases with increasing emission-line luminosity. 

Assuming that most of the optical
continuum of our H{\ts}II and LINER ULIGs
is produced by the underlying stellar population and using
the continuum colors of the stellar cluster models of Jacoby, Hunter,
\& Christian (1984), the underlying optical continuum
in these objects appears to be dominated by a young stellar population 
characterized by a starburst age  of $\sim${\ts}10$^7$ years. 
A similar result was derived from the samples of VKSMS and Armus et al. (1989).
The slightly bluer continuum of the Seyfert 2 ULIGs, though only marginally
significant, may indicate the presence of an additional source of
activity in these objects. 

\subsection{H$\alpha$ Luminosities and Equivalent Widths}

The distributions of H$\alpha$ equivalent widths and reddening-corrected 
H$\alpha$~luminosities are plotted in Figures 16 and 17 for each spectral type.
The median \{log[{\it EW}(H$\alpha$){\ts}\AA], log ({\it L}$_{{\rm H}\alpha}$/{\it L}$_\odot$)$_0$\} are
\{2.1, 9.0\}, \{1.8, 8.7\}, and \{1.9, 9.6\} 
for H{\ts}II galaxies, LINERs, and Seyfert 2 galaxies, respectively.
Both sets of measurements seem to indicate that LINER ULIGs are 
slightly deficient in H$\alpha$ relative to H{\ts}II and Seyfert 2 ULIGs. 
A similar result was found by VKSMS in the lower luminosity objects. 
Rigorous K-S tests indicate, however, that the differences between
the various types of ULIGs are only marginally significant.

A strong correlation is found between H$\alpha$ and 
infrared luminosities (Fig. 18; $P{\rm [null]} = 4\times 10^{-8}$,
$1 \times 10^{-7}$, and 0.005 for the H{\ts}II galaxies, LINERs and
Seyfert 2 galaxies in the combined sample, respectively).
Slightly weaker correlations are observed when the H$\alpha$
luminosities are replaced by the H$\alpha$ equivalent widths
(Fig. 19; $P{\rm [null]} = 3 \times 10^{-6}$, $6 \times 10^{-8}$, and 
0.26, respectively). 
The infrared-to-H$\alpha$~luminosity ratios in Seyfert 2 ULIGs are slightly
smaller than those of H{\ts}II and LINER ULIGs but the difference is not 
statistically significant 
[median log({\it L}$_{\rm ir}$/{\it L}$_{{\rm H}\alpha}$) = 3.11, 3.47, 3.05 for the H{\ts}II,
LINER, and Seyfert 2 ULIGs, respectively]. 
The presence of an additional,
non-stellar source of energy in some of the Seyfert ULIGs may result in 
infrared-to-H$\alpha$ luminosity ratios that differ from the nominal
value of starburst-powered galaxies embedded in dust.

If the H{\ts}II and LINER ULIGs of our sample are purely powered by stars,
both the infrared and H$\alpha$~luminosities can be used to estimate
the rate of star formation in these galaxies.  Using equations (1) and
(2) of VKSMS, prodigious star formations rates of 
$\sim${\ts}250--1,000 $\beta^{-1}$ {\it M}$_\odot$ yr$^{-1}$ are 
derived based on the infrared
luminosities and $\sim${\ts}1--250 $\eta^{-1}$ {\it M}$_\odot$ yr$^{-1}$
based on their {\it nuclear} H$\alpha$ luminosities. The parameters
$\beta$ and $\eta$ in these rates are the fraction of the bolometric
luminosity radiated by stars and the fraction of ionizing photons that
produce observable H$\alpha$ emission, respectively. The difference
between these star formation rates is probably due in most part to the
fact that the H$\alpha$-based rates refer only to the nuclei while the
infrared-based rates take into account the star formation in both the
nuclear and circumnuclear regions.

Little is known about the masses of molecular gas in the ULIGs of our sample.
Therefore, the depletion time scales in these objects are difficult to 
estimate. Sanders, Scoville, \& Soifer (1991) found that the 
infrared luminosity-to-H$_2$ mass ratios for $\sim${\ts}60 BGS galaxies 
increase roughly linearly with increasing $L_{\rm ir}$. 
If this relation also applies to the ULIGs studied in this paper and if the 
observed $L_{\rm ir}$ is powered entirely by star formation, the star
formation rates derived for these galaxies (e.g. Sanders et al. 1991)
will transform the entire interstellar medium into
massive stars in only about 4 $\times$ 10$^7$ {\it L}$_{\rm ir, 12}^{-1}$ yr
(neglecting the mass loss of these stars back into the interstellar
medium and assuming no low-mass stars are formed). 
This timescale is shorter than the lifetime of
the ultraluminous phase estimated
by Carico et al. (1990) from the frequency of double nuclei in these
systems. The time scale for gas depletion may be underestimated if the
ULIGs are anomalously rich in molecular gas (relative to their infrared
luminosity) or if the initial mass function 
is skewed towards massive stars (e.g., Rieke et al. 1993).

\section{Spatial Information}

The previous section presented 
the spectral properties of the nuclear regions
($R < 2$ kpc) of our sample galaxies. This section discusses the behavior of 
these properties as a function of the distance from the nucleus.
The line emission in our sample galaxies often peaks
in the nuclear region. Consequently, 
in order to maximize the signal-to-noise ratio in each radial bin, 
we used extraction windows with sizes that gradually increase with radial
distance from the nucleus.
Two methods of extraction were used. In both cases, 
a central portion of the galaxy was first extracted
from the long-slit spectra, corresponding roughly to the size of the
seeing disk ($R$ = 1 kpc for objects with $z$ $<$ 0.15 and $R$ = 2 kpc
for the other galaxies). To study the effects of aperture 
size on the spectral classification of our sample galaxies, 
extraction windows were used that covered the following radial bins: 
0--1 kpc (exclusively for the objects with $z$ $<$ 0.15), 0--2
kpc, 0--4 kpc, and 0--8 kpc (for the few objects with $z$ $>$
0.15). To determine the radial behavior of 
the spectral parameters, the nuclear portion of the
galaxy was extracted first, followed by extracted portions of the
spectrum on both sides of the nucleus.  The size
of the extraction windows was increased gradually at larger distances from the nucleus.
For this second method, the extraction windows used 
for the low-redshift objects
covered $R$ = 0--1 kpc (data points at $R$ = 0.5 kpc in the figures), 
$R$ = 1--2 kpc (data points at $R$ = 1.5 kpc), and $R$ = 2--4 kpc 
(data points at $R$ = 3 kpc). 
For the objects with $z$ $>$ 0.15, the extraction windows 
covered $R$ = 0--2 kpc (data points at $R$ = 1 kpc in the figures), 
2--4 kpc (data points at $R$ = 3 kpc), 
and $R$ = 4--8 kpc (data points at $R$ = 6 kpc).
 It proved possible to carry out this analysis for
34 objects in our sample. These galaxies are indicated by an asterisk
(*) in Table 2.

A summary of the results is presented in Table 7 and in Figures 20--22.
This last figure deserves further explanations:
to attempt to detect any systematic differences between the various
spectral types, we plotted in Figure 22 the mean spectral parameters
of each spectral type, with the error bars in this figure indicating
the standard deviations from the means in each radial bin. 
For this exercise, the mean of the spectral parameters were selected for the 
following radial bins: 0--1 kpc, 1--2 kpc, 2--4 kpc, and 4--8 kpc. 
This technique therefore results in the loss of some data 
because it does not take into account the nuclear
parameters of the high-redshift galaxies extracted over $R$ = 0--2 kpc.
Moreover, one should note that the radial bins 0--1 kpc and 4--8 kpc 
are covered only by the low and high-redshift objects, respectively.
Nevertheless, we believe that Figure 22 is a powerful tool for summarizing our
results. 

\subsection{Spectral Classification}
 
Figure 20 shows the variations of the line ratios as 
a function of aperture size. The spectral types 
of $\sim${\ts}20\% of the ULIGs are found to depend on the size of the 
aperture (6, 7, and 5 objects 
in the diagrams involving [O{\ts}I]/H$\alpha$, [S{\ts}II]/H$\alpha$, and 
[N{\ts}II]/H$\alpha$ plots, respectively). A similar fraction ($\sim${\ts}25\%) 
of the VKSMS sample of LIGs was found to change spectral type
with aperture size. Once again, this result emphasizes the fact
that a constant linear-size aperture is crucial when comparing 
the nuclear emission-line spectra of LIGs having 
different redshifts. 

Dramatic radial variations of the line ratios are observed in several of the
galaxies (Fig. 21). In particular, an important fraction of the galaxies 
with nuclear H{\ts}II characteristics harbor distinct AGN-like line ratios
(generally LINER-like but in some cases even Seyfert-like) 
at larger distances from the nucleus
(5, 8, and 3 objects in the [O{\ts}I]/H$\alpha$, [S{\ts}II]/H$\alpha$, and
[N{\ts}II]/H$\alpha$ plots, respectively). 
This behavior was also observed in a few objects from the sample of
VKSMS. This result was interpreted by VKSMS 
in the context of supernova-driven wind
models in which the circumnuclear gas is
collisionally ionized by strong shocks caused by the interaction of 
the outflowing nuclear gas with the ambient material (e.g., M{\ts}82;
Heckman, Armus, \& Miley 1990). 
The broad [O{\ts}III] line widths observed outside the nuclei of
many H{\ts}II ULIGs (Fig. 22) brings support to this idea.
Shocks may also explain the Seyfert-like line ratios observed in a few
cases (e.g., Binette, Dopita, \& Tuohy 1985; Dopita \& Sutherland 1995). 

A careful inspection of Figure 21 reveals that 
the level of ionization of most galaxies with nuclear
Seyfert 2 spectra tends to
decrease at larger radii. The [O{\ts}III]/H$\beta$ ratios of these objects
are often lower outside of the nucleus and [O{\ts}I]/H$\alpha$,
[S{\ts}II]/H$\alpha$, and [N{\ts}II]/H$\alpha$ often increase with radius. 
A direct comparison of these results with the line ratios predicted by
photoionization models (e.g., Ferland \& Netzer  1983; VO87) suggests
that the ionization parameter ($\equiv$ density of ionizing
photons/hydrogen density) is decreasing away from the nucleus of
these galaxies, as expected if the source of ionization is a point
source such as an AGN and if the density profile is not a steeply
decreasing function of distance from the nucleus. 
Circumnuclear starbursts may also contribute
in some cases to dilute the AGN emission and decrease the degree of
ionization of the gas surrounding the AGN. 

\subsection{Reddening}

The radial variations of {\it E(B--V)} derived from H$\alpha$/H$\beta$
are presented in Figure 22a.  For simplicity, it was assumed that the
intrinsic Balmer decrements outside of the nucleus is the same as in
the nucleus [i.e., it was assumed (H$\alpha$/H$\beta$)$_0$ = 2.85 in
and out of the H{\ts}II nuclei and 3.1 in and out of the AGN-like nuclei].
No significant differences are observed between the mean reddening
profiles of the various spectral types (Fig. 22a). The color excess
shows a general tendency to decrease with radius. However, there is
substantial scatter around the mean profile of each spectral type.  In
at least 4 and possibly 7 galaxies (FSC{\ts}10494+4424, FSC{\ts}14378-3615,
FSC{\ts}16333+4630, FSC{\ts}17208-0014; FSC{\ts}11582+3020, FSC{\ts}16468+5200, FSC{\ts}17044+6720), the
amount of reddening appears to peak outside of the nucleus. Six of
these galaxies present a LINER-like nuclear spectrum.  This fraction
of galaxies with inverted reddening profiles is qualitatively similar
to that found by VKSMS in the LIGs: 2 LINERs and 1 Seyfert 2 galaxy
among the 23 LIGs.  It is not clear at present whether inverted
reddening profiles reflect the actual dust distribution in these
objects or whether they are caused by complex optical depth effects.
High-resolution ($\lesssim$ 1\arcsec) observations of our 1-Jy
galaxies at millimeter wavelengths could prove useful for determining
how the molecular gas is distributed in these ULIGs.

\subsection{Line Widths}

The radial variations of the [O{\ts}III] line widths are presented in
Figure 22b. We note that the Seyfert 2 ULIGs present
[O{\ts}III] profiles that are broader than those of H{\ts}II and LINER
ULIGs not only in the nucleus (\S 4.4) but also out to radii of $\sim${\ts}2--3 kpc. 
The large line widths in the circumnuclear regions of the
Seyfert 2 ULIGs suggest that the gas near the nucleus
is feeling the dynamical effects of the AGN presumed to exist in these
objects. This process is known to take place in a number of optically-selected
Seyfert galaxies (e.g., Whittle 1994). 

It is harder to explain the apparent tendency for the [O{\ts}III] profiles
in H{\ts}II and LINER ULIGs to be broader {\it outside} of the
nucleus. One possibility --- proposed by VKSMS to explain their own
data --- is that the broad line profiles outside of the nuclei of these
objects are produced by (unresolved) line splitting where two or more
kinematic components with different velocities are
present. Extra-nuclear line splitting due to nuclear outflows is a
relatively common feature in the handful of LIGs for which
high-resolution kinematic data exist in the literature (e.g., Bland \&
Tully 1988; Heckman, Armus, \& Miley 1990; Veilleux et al. 1994 and
references therein). It is possible that this process is taking place
in some of our sample galaxies (see \S 5.1), but this process may have
difficulties explaining the fact that the line widths of an important
fraction of the H{\ts}II and LINER galaxies are {\it monotonically}
increasing outward from the nucleus.  Another possible source of
broadening is galactic rotation.  The importance of this effect is a
complex function of the rotation curve, the distribution of the line
emission, and the size of the aperture. Under certain conditions, the
radially increasing size of our extraction apertures may indeed result
in line smearing/broadening that could reproduce the
observations. Unfortunately, the spectral resolution of the present
data is not sufficient to answer this question. Finally, we cannot
completely rule out the possibility that the measurements obtained
outside of the nuclei are affected by numerical broadening associated
with Poisson noise (noisier signals tend to pull in more of the
emission in the wings away from the peak and so broaden the fitted
line profile). Data of higher signal-to-noise ratios and spectral
resolution would settle this controversy unambiguously.

\subsection{Stellar Absorption Features, Continuum Colors and H$\alpha$
Luminosities}

The radial behavior of the H$\beta$ and Mg Ib equivalent widths are
plotted in Figures 22$c$ and 22$d$. The strength of both features is 
relatively constant 
with radius. 
The Mg{\ts}Ib  feature {\it at all radii} is generally weaker than
that of non-active spiral galaxies 
[{\it EW}(Mg Ib) = 3.5--5.0{\ts}\AA: Keel 1983; Stauffer 1982; Heckman, Balick,
\& Crane 1980]. The strength of the H$\beta$
feature, on the other hand, covers a range that is similar to that of
non-active spiral galaxies ($\sim${\ts}1.4{\ts}\AA: Keel 1983). It is hard to
simply invoke dilution effects by a featureless continuum from
hot stars to explain both sets of results simultaneously and at all
radii. A natural explanation of these radial behaviors 
is that the old (Mg{\ts}Ib-producing) 
stellar population present in non-active galaxies 
is less prominent in ULIGs. Dilution effects probably become an
important factor in the nucleus, however (especially in Seyfert 2 galaxies;
see \S 4.5 and below). 

Figure 22$e$ presents the radial variations of the 
reddening-corrected continuum colors. 
Irrespective of spectral types, the median values of (C6563/C4861)$_0$
in ULIGs are $\sim${\ts}0.4 within $R$ = 0--3 kpc.
The estimated age of the underlying stellar population with these
continuum colors is $\sim${\ts}10$^7$ years (Jacoby et al. 1984;
VKSMS).

None of the apparent differences between spectral types (Fig.  22$c$,
22$d$, and 22$e$) are statistically significant. However, the slightly
smaller {\it nuclear} values of {\it EW}(H$\beta$), {\it EW}(Mg{\ts}Ib),
and (C6563/C4861)$_0$ in the Seyfert 2 ULIGs are consistent with the
presence of an additional source of activity in the nuclei of these
objects.  The nuclear peak of the H$\alpha$ emission in Seyfert
galaxies (Fig. 22$f$) --- in spite of the fact that the size of the
extraction windows increases with radius --- confirms that the activity
in these objects is more highly concentrated in the nucleus than in
any other classes of ULIGs.  The peak is slighty less pronounced when
{\it L}$_{{\rm H}\alpha}$ is replaced by {\it EW}(H$\alpha$) (see
Fig. 22$g$).

\section{Summary}

An optical spectroscopic study of the nuclear regions 
of an unbiased subsample of 45 ultraluminous infrared galaxies from Paper I, 
has been completed. The spectral properties of the 
circumnuclear regions in 34 of these objects were also examined.
These data were then combined with our results from a previous study of 
primarily lower luminosity infrared galaxies (VKSMS) to examine 
the spectra properties of LIGs over the range 
$L_{\rm ir} \approx 10^{10.5}-10^{12.9}\ L_\odot$. 
The main conclusions of this analysis are the following:

\begin{enumerate}

\item
The fraction of LIGs with Seyfert
characteristics continues to increase with increasing $L_{\rm ir}$.  
For $L_{\rm ir} > 10^{12.3}\ L_\odot$, about 45\% 
of the ULIGs (9/20 objects) are classified as Seyfert 1s
or 2s.

\item
Many properties of the Seyfert galaxies in our sample suggest the 
presence of a nuclear source of activity in these objects 
that is absent or not visible in H{\ts}II
galaxies or LINERs. The distribution of the H$\alpha$ emission
in Seyfert galaxies is more centrally peaked than in H{\ts}II galaxies and
LINERs. 
The nuclei of Seyfert galaxies present a slightly bluer continuum and weaker
stellar features. The line widths within $R \sim 2-3$ kpc of the nuclei
of Seyfert galaxies are broader on average than those measured in the H{\ts}II
and LINER ULIGs. The variations of the line
ratios in Seyfert galaxies imply that the ionization parameter is
decreasing away from the nucleus of many of these galaxies, as
expected if the source of ionization is a point source like an AGN. 

\item
In contrast, the LINER
ULIGs share more often than not the properties of H{\ts}II galaxies, and 
therefore may not be genuine AGN. In some cases, the
radial variations of the line ratios suggest that the
extranuclear LINER-like emission is produced through shocks caused by the
interaction of starburst-driven outflows with the ambient material. 
Note that this statement only refers to the infrared-selected LINERs of
our sample, and may not necessarily apply to optically-selected LINERs. 

\item
The continuum colors and strengths of the stellar H$\beta$ and Mg{\ts}Ib
features in and out of the nuclei of ULIGs indicate that
a young ($\sim${\ts}10$^7$ year) stellar population is present at all radii
and overwhelms
the emission from the old, underlying stellar population in these
objects. Star formation rates of $\sim${\ts}250--1,000
{\it M}$_\odot$ yr$^{-1}$ and $\sim${\ts}1--250 {\it M}$_\odot$ yr$^{-1}$ are
derived from the {\it total} infrared luminosities and {\it nuclear} H$\alpha$
luminosities, respectively. These prodigious star formation rates will
transform the entire interstellar medium into massive stars in only a
few $\times 10^7$ yr unless these galaxies
are anomalously rich in molecular gas (relative to their infrared
luminosity) or if the initial mass function 
is skewed towards massive stars.

\item
Based on the color excesses derived from the H$\alpha$/H$\beta$
emission-line ratios, the importance of dust in the nuclei of ULIGs
appears comparable to that measured in the {\it IRAS} galaxies of
lower infrared luminosities.  However, in constrast to what was found
in low-luminosity infrared galaxies, the color excess in the nuclei of
ULIGs does not seem to depend on spectral types.  The reddening in
both the high and low luminosity objects is generally observed to
decrease with distance from the nucleus.  Clear cases of inverted
reddening profiles are observed in less than 1/5 of the ULIGs in our
sample, nearly all of which are optically classified as LINERs.  Complex
optical depth effects may distort our view of the dust distribution in
these objects, and make the interpretation of these inverted
reddening profiles difficult.  High-resolution observations of these
objects at millimeter/submillimeter wavelengths would clarify this
issue.

Optical spectra of the remaining objects in the 1-Jy sample have
been collected and are presently being analyzed (Veilleux et al. 
1997a).  An analysis of near-infrared spectra on a
representative subset of 25 objects from the 1-Jy sample was also
recently presented (Veilleux, Sanders, \& Kim 1997b).  These data will
be combined together into a large optical/near-infrared data base that
will allow us to make firmer statements on the statistical
significance of the results discussed in the present paper.

\end{enumerate}

\acknowledgments
The IRAF procedure ``specfit'' was developed and kindly provided by
Gerard A. Kriss with funding by NASA to the Hopkins Ultraviolet
Telescope project. 
This research was supported in part by NASA grants NAG5-1251 and NAG5-1741 
to the
University of Hawaii (D. B. S. and D.-C. K.). S. V. gratefully
acknowledges the financial support of NASA through grant number
HF-1039.01-92A awarded by the Space Telescope Science Institute which
is operated by the AURA, Inc. for NASA under contract No. NAS5-26555.
This research has made use of the NASA/IPAC Extragalactic Database
(NED) which is operated by the Jet Propulsion Laboratory, California
Institute of Technology, under contract with NASA.

\clearpage

\centerline{FIGURE CAPTIONS}

\vskip 0.3in

\figcaption[Fig1.eps]{Optical spectra of the ULIGs in 
our sample --- $f_\lambda$ is plotted vs $\lambda_{\rm observed}$. 
The units of the vertical axis are 10$^{-15}$ erg s$^{-1}$ cm$^{-2}$ 
\AA$^{-1}$ while the wavelength scale is in{\ts}\AA.
}

\figcaption[Fig2.eps]{Dereddened flux ratios. 
The H{\ts}II region-like galaxies (H) are located to the left of the solid
curve while AGN-like objects are located on the right.
AGN-like objects were further classificated as Seyfert-like or LINER-like 
depending on whether or not [O{\ts}III]$\lambda$5007/H$\beta\ \geq$ 3 (indicated 
by a solid horizontal segment).
Seyfert 1 galaxies are not shown in the diagrams.
}

\figcaption[Fig3.eps]{Summary of the results of the spectral classification as a
function of the infrared luminosity. The results from the present
paper were combined with those obtained by Veilleux et al. (1995) on
LIGs in the BGS. 
The fraction of Seyferts increases with infrared luminosity.
}

\figcaption[Fig3.eps]{Distribution of the color excesses as a function of 
spectral types for the ULIGs in the 
present sample (solid line) and the
BGS LIGs of Veilleux et al.  (1995; dashed line). 
Both classes of objects present large color excesses. 
The differences between the various types of
ULIGs are not significant. 
}

\figcaption[Fig5.eps]{Color excesses as a function of the 
equivalent widths of Na{\ts}Id $\lambda$5892 for each spectral type. 
The stars are the H{\ts}II galaxies, the open circles are the LINERs,
and the filled circles are the Seyfert 2 galaxies. The large symbols
are the ULIGs from the present sample and the small symbols are the
BGS LIGs from Veilleux et al.  (1995). A significant correlation is
observed among H{\ts}II galaxies.
}

\figcaption[Fig6.eps]{Color excesses as a function of observed continuum colors
for each spectral type. The meaning of the symbols is the same as in Fig. 5.
A correlation is observed among H{\ts}II galaxies and LINERs.
}

\figcaption[Fig7.eps]{Distribution of the electron density as a function of 
the spectral types of the ULIGs in the present sample
(solid line) and the
BGS LIGs of Veilleux et al.  (1995; dashed line). No luminosity or spectral
type dependence is observed.
}

\figcaption[Fig8.eps]{Distribution of the [O{\ts}III] $\lambda$5007 line widths 
as a function of the spectral types of the ULIGs in the 
present sample (solid line) and the
BGS LIGs of Veilleux et al.  (1995; dashed line). 
The median line widths of H{\ts}II and
LINER ULIGs are somewhat smaller than those of Seyfert 2 ULIGs, but the
difference is not significant.
}

\figcaption[Fig9.eps]{[O{\ts}III] $\lambda$5007 line widths as a function of the infrared
luminosities. The meaning of the symbols is the same as in Fig. 5. The
objects with large [O{\ts}III] line widths generally have large infrared
luminosities and many of them have Seyfert characteristics. 
}

\figcaption[Fig10.eps]{[O{\ts}III] $\lambda$5007 
line widths as a function of the flux density ratios
${\it f}_{25}/{\it f}_{60}$ for each spectral type. 
The meaning of the symbols is the same as in Fig. 5. A correlation is
observed among Seyfert 2 galaxies. 
}

\figcaption[Fig11.eps]{Distribution of the equivalent widths of
H$\beta$ in absorption as 
a function of the spectral types of the ULIGs in the present sample
(solid line) and the BGS LIGs of Veilleux et al.  (1995; dashed line).
The equivalent widths of the ULIGs are on average smaller than those
of the lower-luminosity objects. Seyfert ULIGs may also have smaller
equivalent widths than the other two classes of ULIGs, but the
difference is not significant.  }

\figcaption[Fig12.eps]{Distribution of the equivalent widths of Mg{\ts}Ib $\lambda$5176 
as a function of the spectral types of the ULIGs in the 
present sample (solid line) and the
BGS LIGs of Veilleux et al.  (1995; dashed line). 
No significant differences between spectral types are detected among ULIGs.
However, these equivalent widths are significantly smaller than those
of non-active spiral galaxies (not shown here).
}

\figcaption[Fig13.eps]{Distribution of the dereddened continuum colors, (C6563/C4861)$_0$,
as a function of the spectral types of the ULIGs in the 
present sample (solid line) and the
BGS LIGs of Veilleux et al.  (1995; dashed line). 
The nuclear continuum of Seyfert ULIGs
appears bluer on average than those of H{\ts}II and LINER ULIGs, but
these differences are not significant.
}

\figcaption[Fig14.eps]{Dereddened continuum colors as a function of the dereddened
H$\alpha$ luminosities. The meaning of the symbols is the same as in
Fig. 5. A strong anti-correlation is observed.
}

\figcaption[Fig15.eps]{Dereddened continuum colors as a function of the infrared
luminosities. The meaning of the symbols is the same as in
Fig. 5. 
}

\figcaption[Fig16.eps]{Distribution of the equivalent widths of H$\alpha$ in
emission as a function of the spectral types of the ULIGs in the 
present sample (solid line) and the
BGS LIGs of Veilleux et al.  (1995; dashed line). 
The average equivalent width of
the LINERs is marginally smaller than that of the H{\ts}II galaxies, both
at high and low infrared luminosities.
}

\figcaption[Fig17.eps]{Distribution of the reddening-corrected 
H$\alpha$ luminosities as a function of the spectral types of the
ULIGs in the present sample (solid line) and the
BGS LIGs of Veilleux et al. (1995; dashed line). 
The LINERs are somewhat
underluminous relative to Seyfert 2 galaxies, both at high and low
infrared luminosities. 
}

\figcaption[Fig18.eps]{H$\alpha$ luminosities as a function of infrared
luminosities. The meaning of the symbols is the same as in
Fig. 5. Significant correlations are observed for all three spectral types. 
}

\figcaption[Fig19.eps]{Equivalent widths of H$\alpha$ as a function of
infrared luminosities. The meaning of the symbols is the same as in
Fig. 5. Significant correlations are observed among H{\ts}II galaxies and
LINERs.  }

\figcaption[Fig20.eps]{Dereddened line ratios as a function of the size of the
apertures. In all three panels, the asterisks (*) mark the nuclear values.
The size of the extraction aperture increases along each segment (see
introduction of \S 5). 
}

\figcaption[Fig21.eps]{Dereddened line ratios as a function of the distance from the nucleus.
In all three panels, the asterisks (*) mark the nuclear values. The
mid-distance of the extraction aperture increases along each segment (see
introduction of \S 5). 
}

\figcaption[Fig22.eps]{Radial profiles of the mean values of (a) the
color excesses, (b) the [O{\ts}III] $\lambda$5007 line widths, (c) the
equivalent widths of H$\beta$ in absorption, (d) the
equivalent widths of Mg{\ts}Ib, (e) the dereddened continuum colors, (f)
the H$\alpha$ luminosities, and (g) the equivalent widths
of the H$\alpha$ emission.  Stars, open circles, and filled circles
represent data points from the H{\ts}II, LINERs and Seyfert 2 ULIGs,
respectively. The error bars in these figures indicate the standard
deviation from the means in each radial bin. }

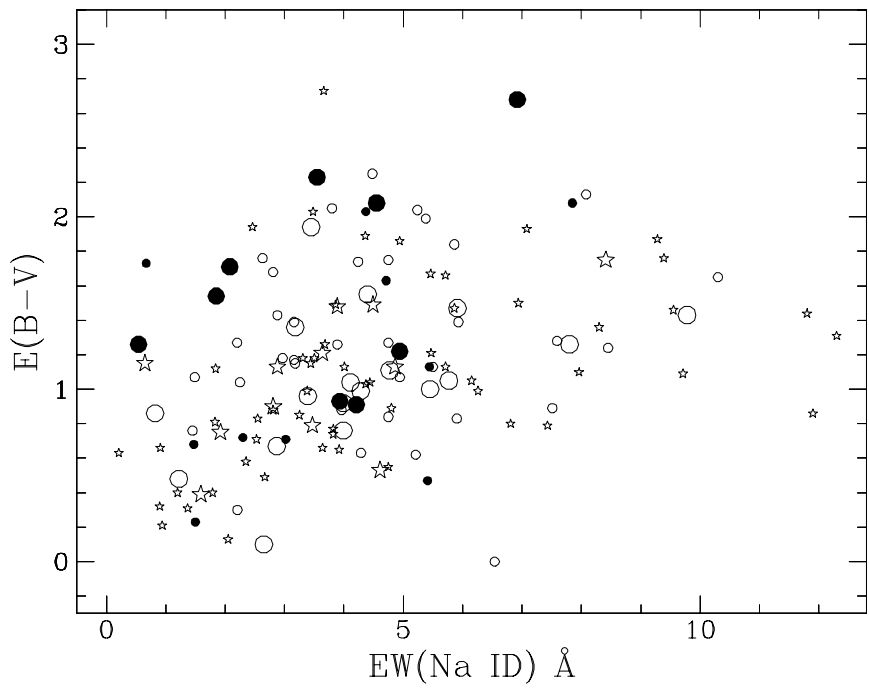
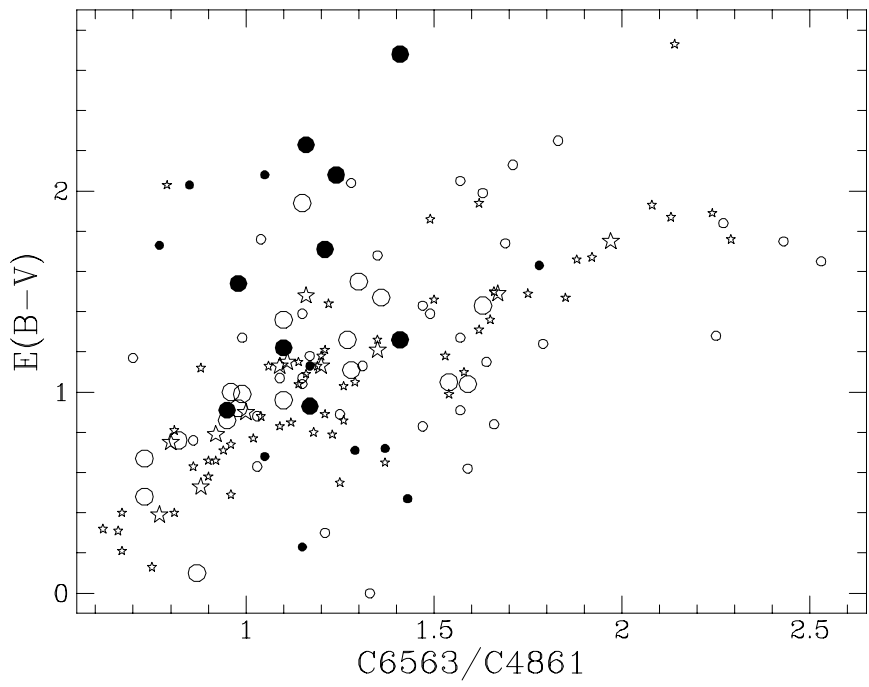
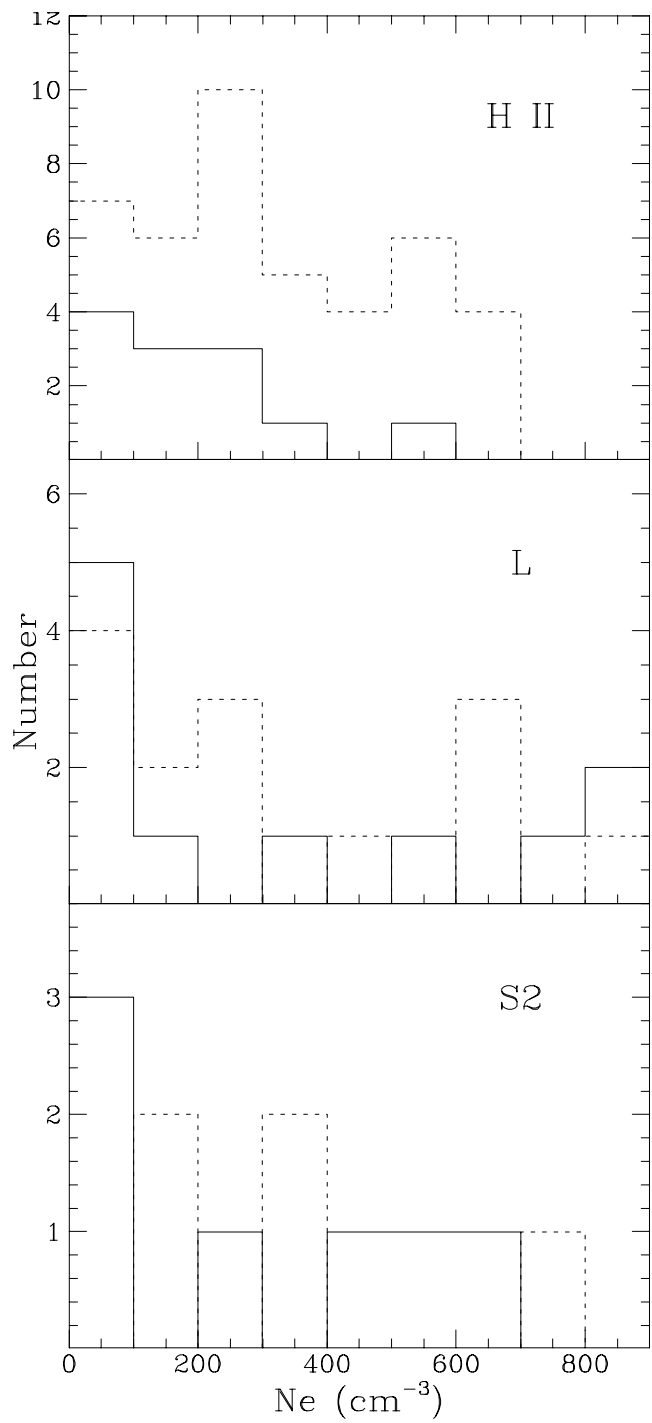
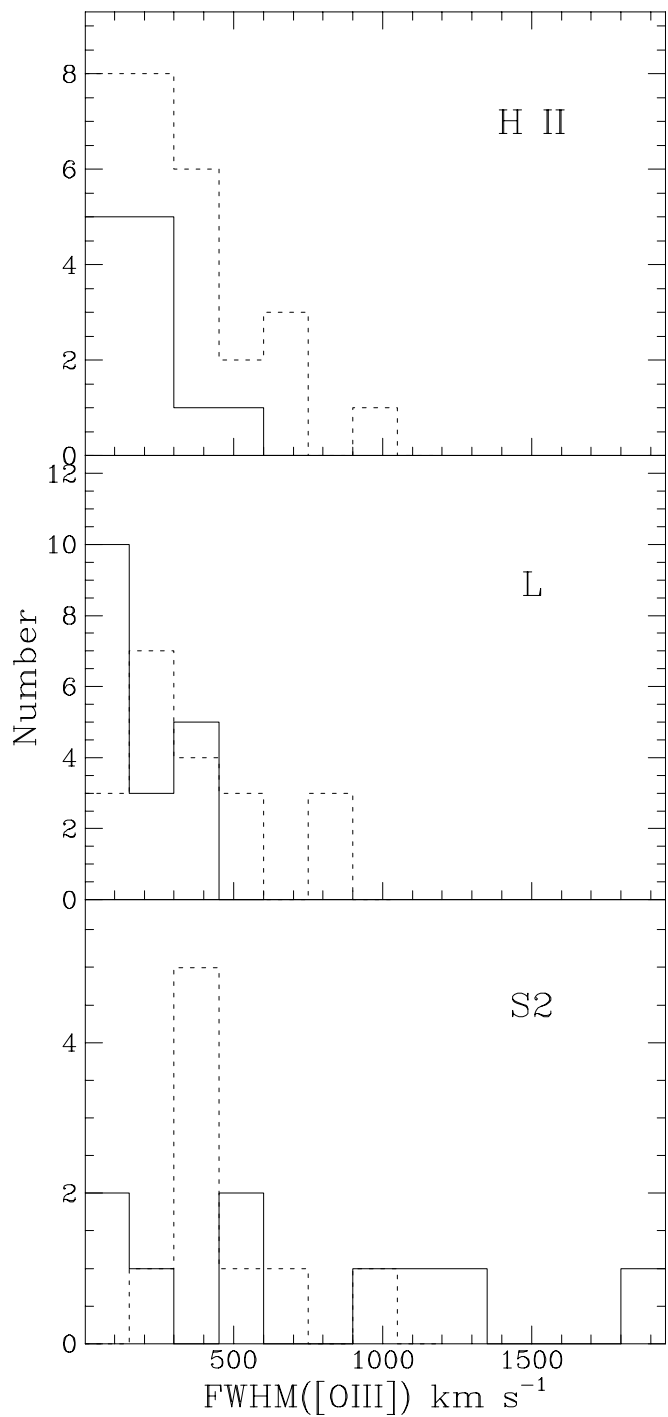
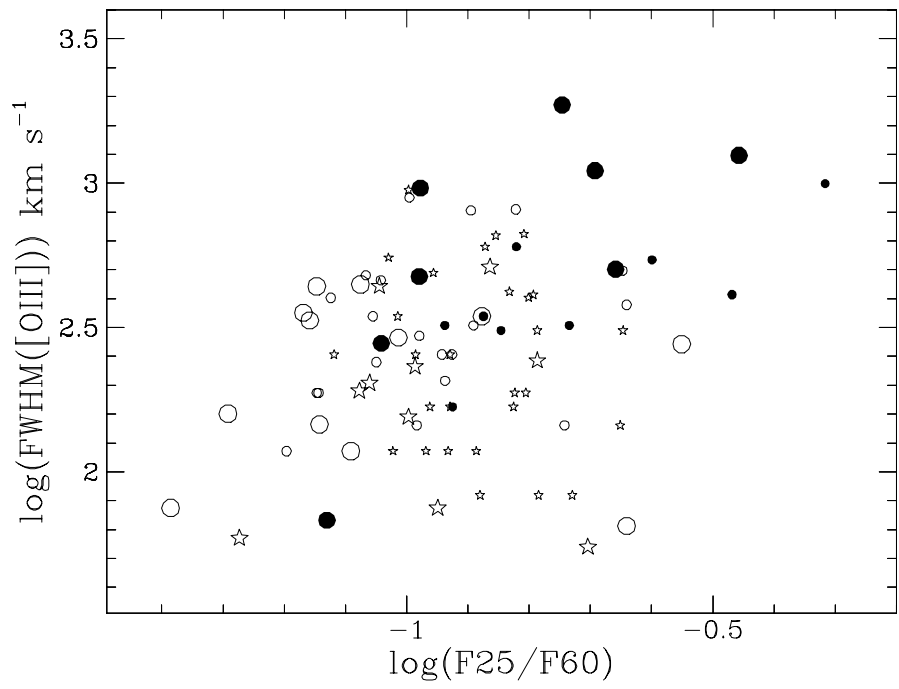
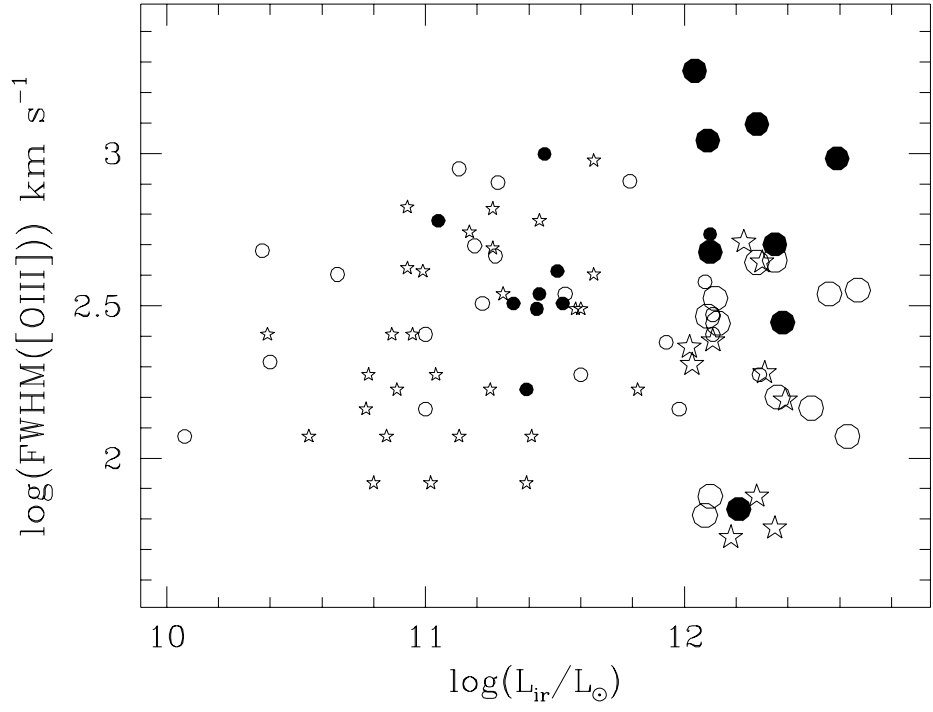
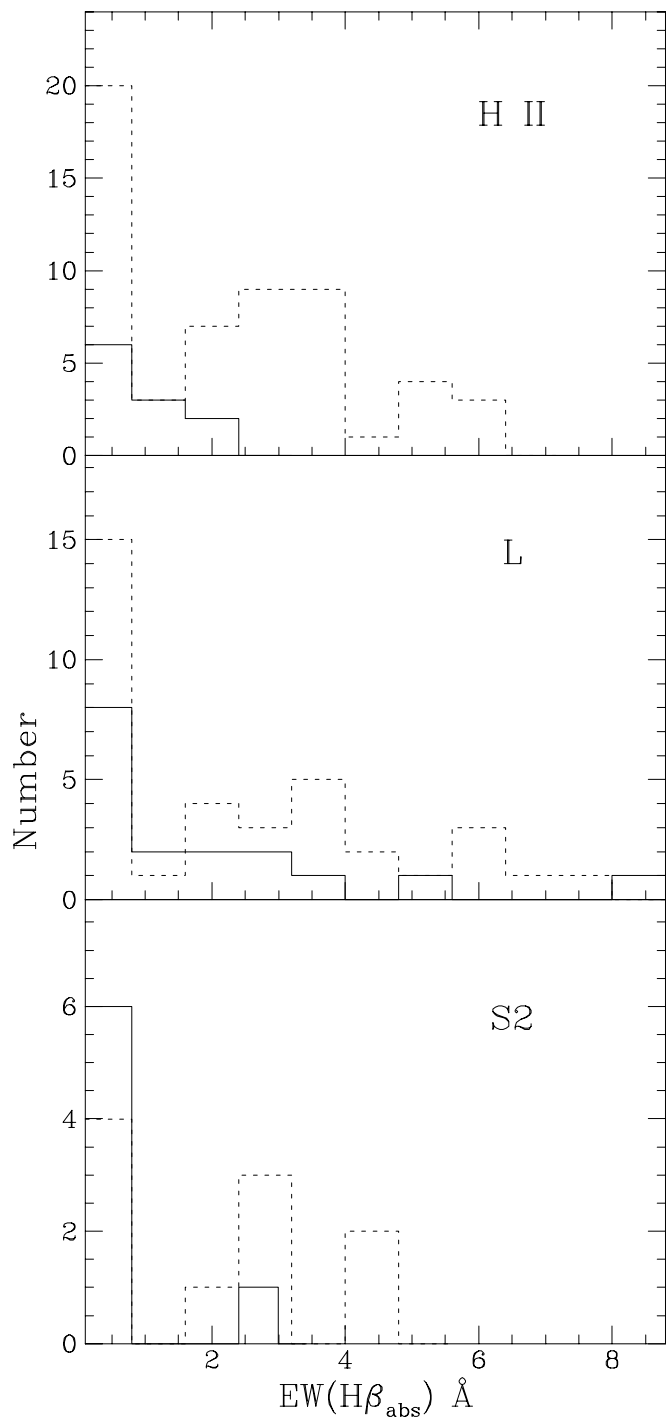
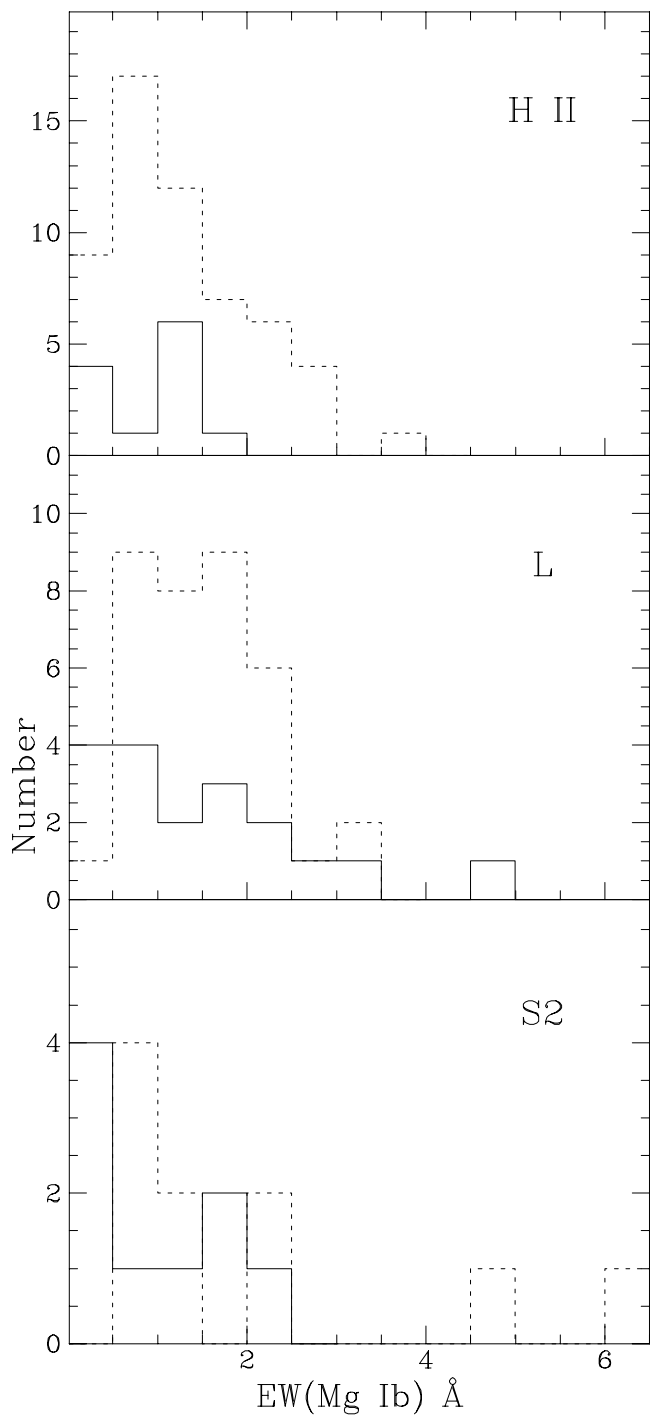
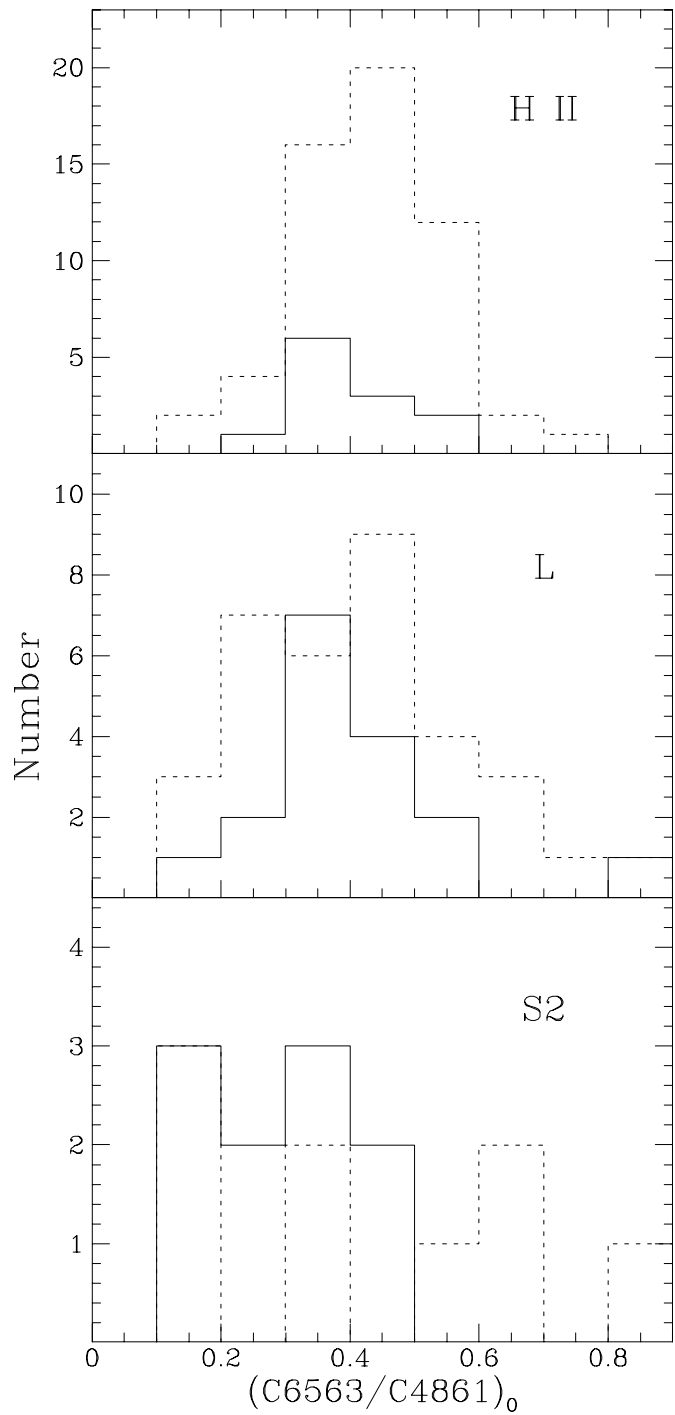
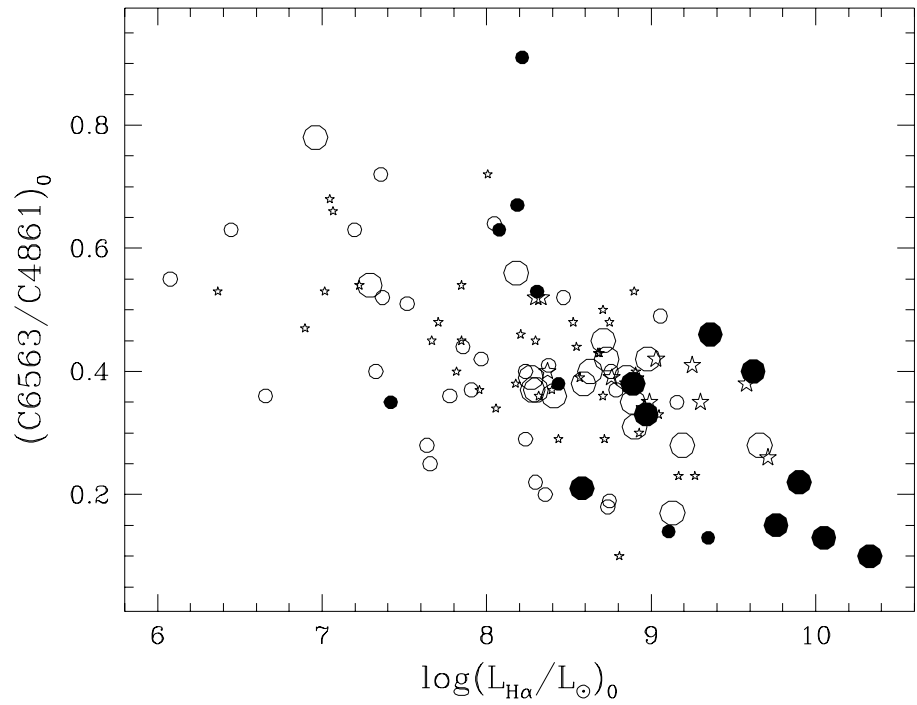
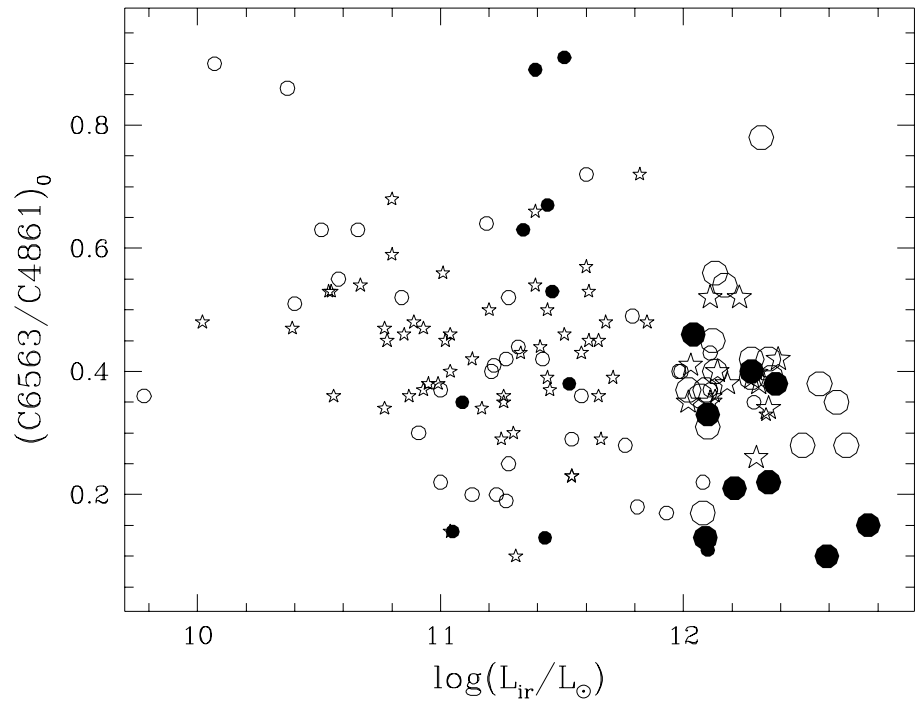
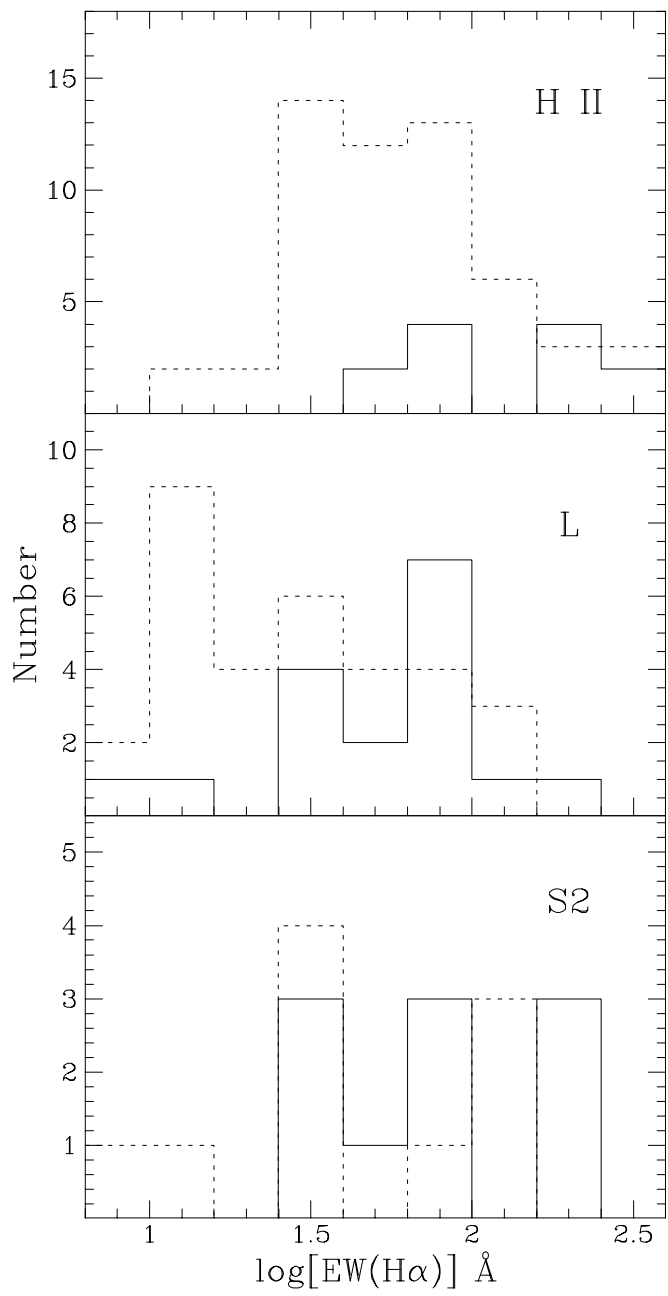
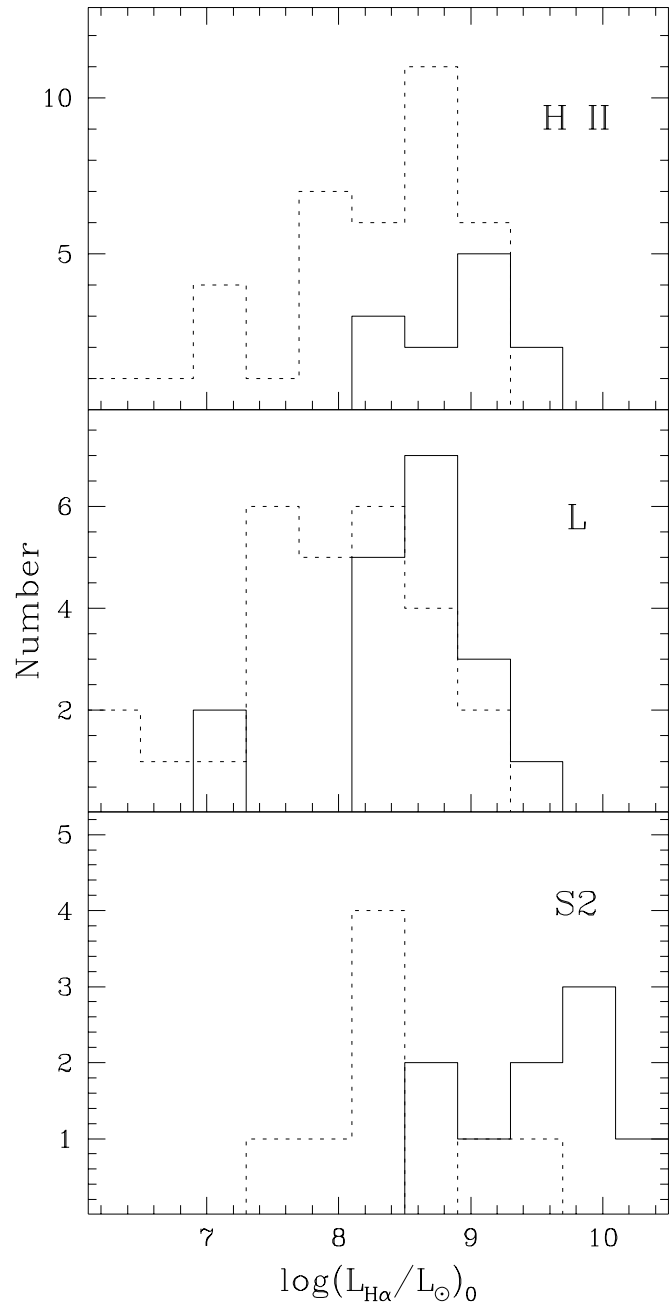
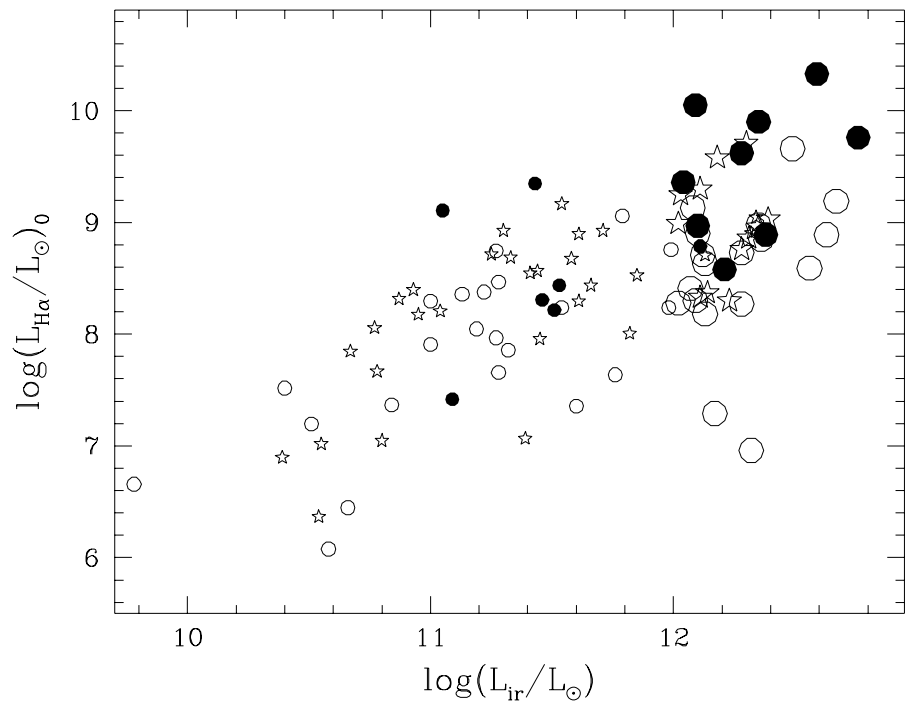
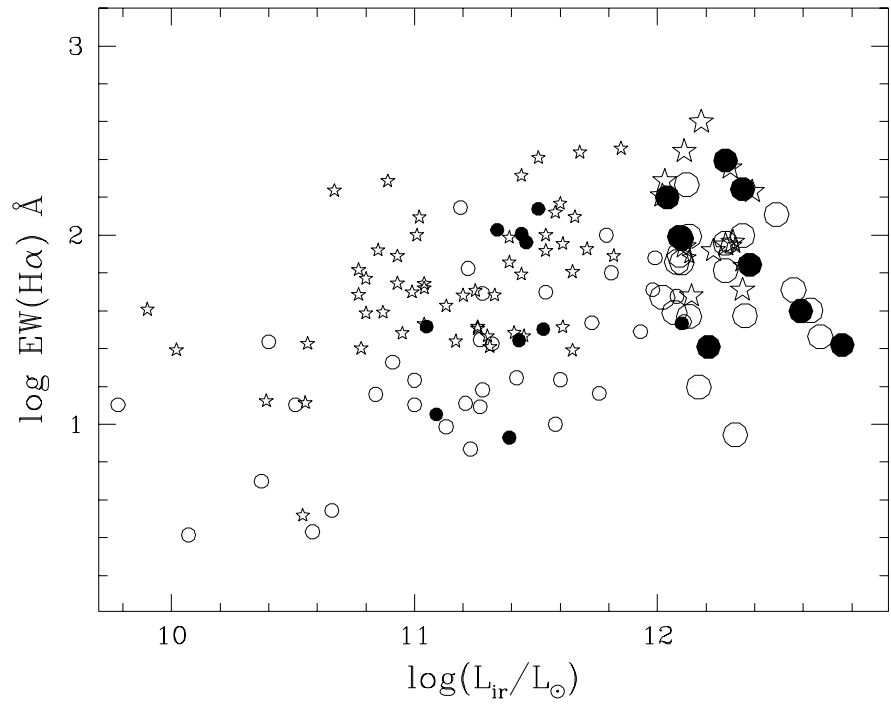
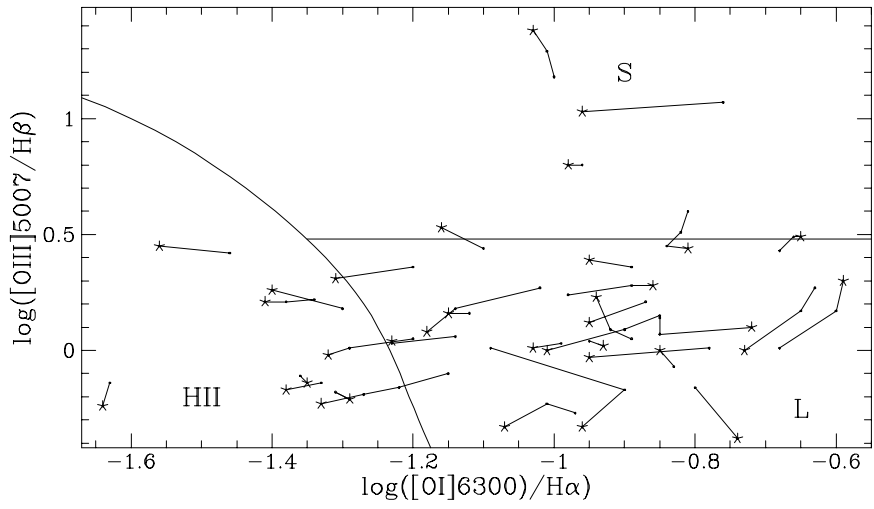
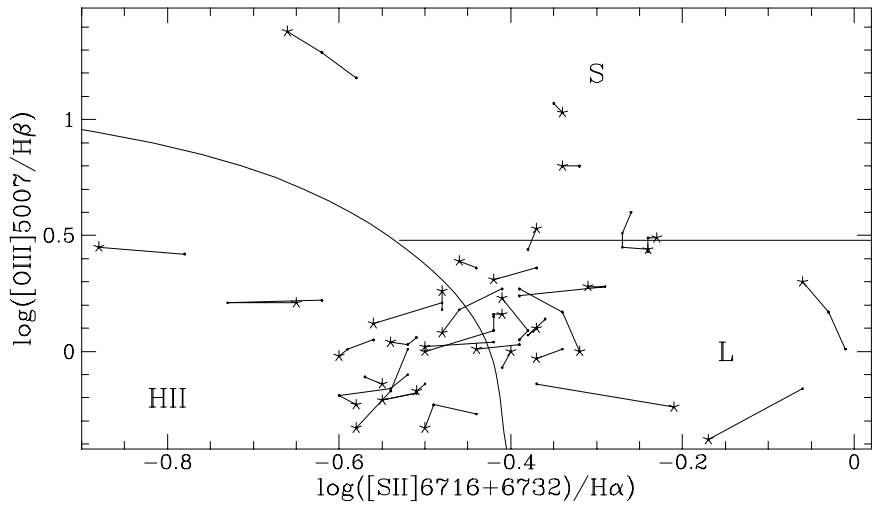
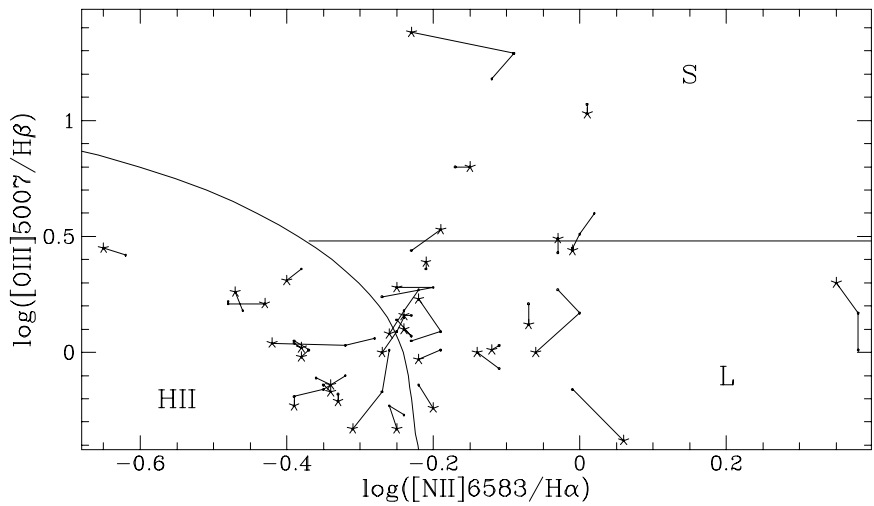
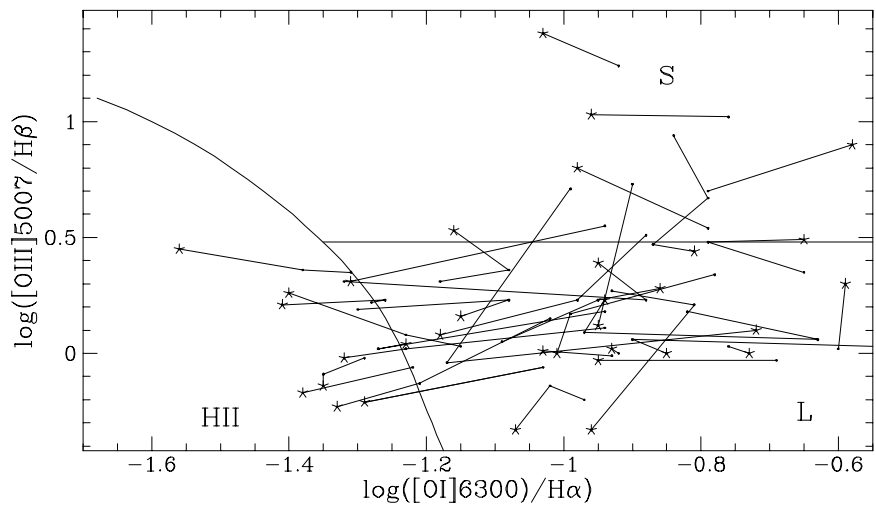
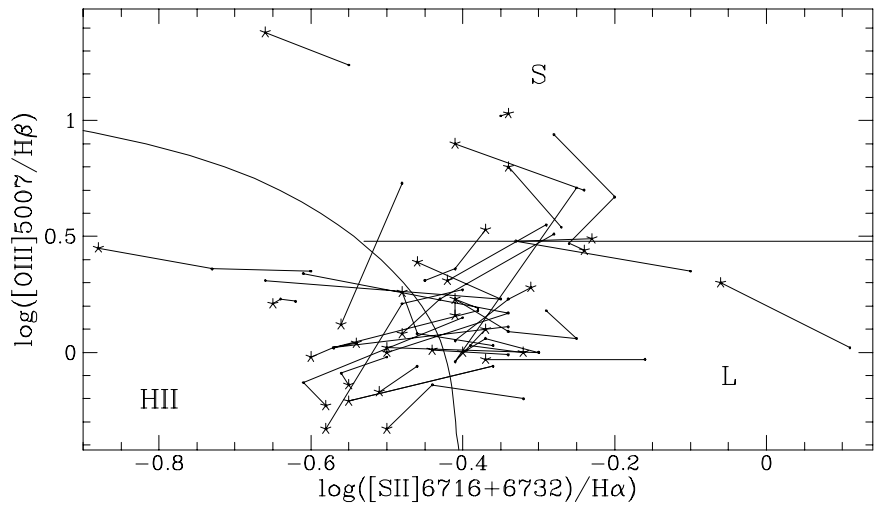
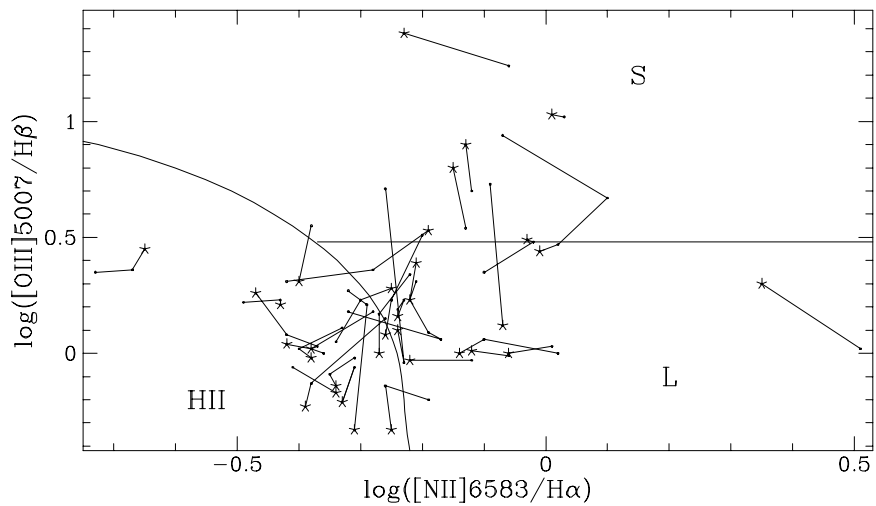
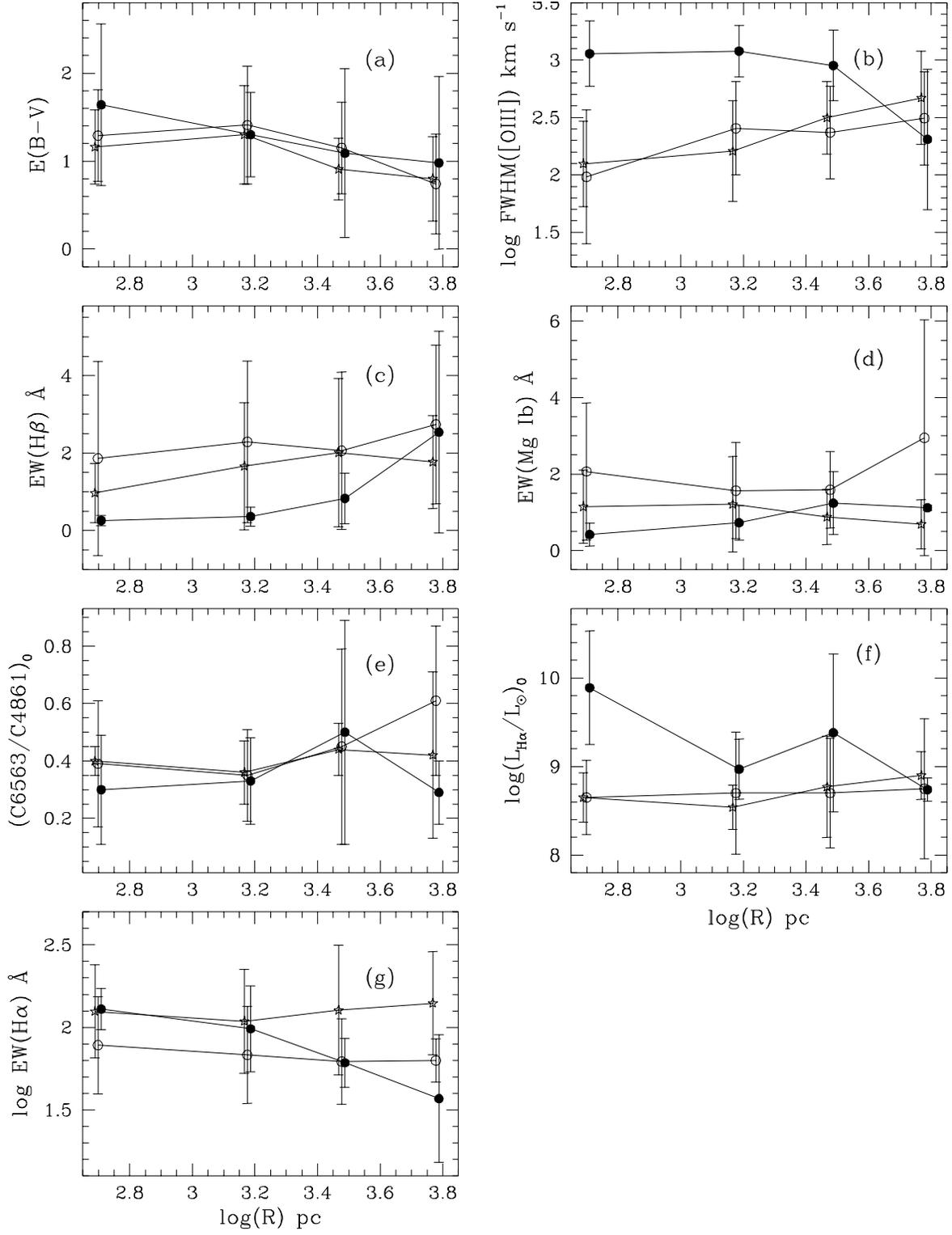
\end{document}


\setcounter{page}{48}
\begin{deluxetable}{lccccc}
\tablenum{1}
\tablewidth{0pt}
\tablecaption{Journal of Observations}
\tablehead{
\colhead{Obs. Date} &
\multicolumn{1}{c}{Grating} &
\multicolumn{1}{c}{Spectral Range} &
\multicolumn{1}{c}{Resolution} &
\multicolumn{1}{c}{Seeing} &
\colhead{CCD} \nl
\colhead{} &
\multicolumn{1}{c}{(l/mm)} &
\multicolumn{1}{c}{(\AA)} &
\multicolumn{1}{c}{(\AA)} &
\multicolumn{1}{c}{($^{\prime\prime}$)} & 
\colhead{} 
}
\startdata
1992 Jul 5&600&4730 $-$ 7740&8.3&1.1--1.3&Tek 1024$^2$ \nl
1993 Mar 30-31&600&5670 $-$ 8770&8.3&1.2--2.0&Tek 2048$^2$ \nl
1993 May 14-15&600&4920 $-$ 8270&8.3&1.2--1.8&Tek 2048$^2$ \nl
\enddata
\end{deluxetable}

\clearpage

\newcommand{\ts}{\thinspace}
\begin{deluxetable}{llcl}
\small
\tablenum{2}
\tablewidth{4in}
\tablecaption{Spectroscopic 1 Jy Survey}
\tablehead{
\colhead{IRAS FSC} &
\multicolumn{1}{c}{$z$} &
\multicolumn{1}{c}{log$\left( L_{\rm ir} \over L_\odot \right)$} &
\colhead{Type} \nl
\colhead{(1)} &
\multicolumn{1}{c}{(2)} &
\multicolumn{1}{c}{(3)} &
\colhead{(4)} 
}
\startdata
\sidehead{\hskip 1truein New Data for 1 Jy Sample (39 objects)\tablenotemark{a}}
07599+6508&   0.149 (1)&12.46&   Q,S1 (1,5,6) \nl
08201+2801$^*$&   0.168 (1)&12.23&   H (1) \nl
08572+3915$^*$&   0.058 (1)&12.11&   L (1,5) \nl
10091+4704&   0.246 (1)&12.67&   L (1) \nl
10494+4424$^*$&   0.092 (1)&12.13&   L (1) \nl
11029+3130&   0.199 (1)&12.32&   L (1) \nl
11119+3257&   0.189 (1)&12.58&   S1 (1) \nl
11223$-$1244$^*$&   0.199 (1)&12.59&   S2 (1) \nl
11506+1331$^*$&   0.127 (1)&12.28&   H (1) \nl
11582+3020$^*$&   0.223 (1)&12.56&   L (1) \nl
 & & & \nl
12072$-$0444$^*$&   0.129 (1)&12.35&   S2 (1,5) \nl
12112+0305$^*$&   0.073 (1)&12.28&   L (1) \nl
12540+5708&   0.042 (1)&12.50&   Q,S1 (1,5,6) \nl
13218+0552&   0.205 (1)&12.63&   S1 (1,4) \nl
13342+3932&   0.179 (1)&12.37&   S1 (1) \nl
13428+5608$^*$&   0.037 (1)&12.10&   S2 (1,5)\tablenotemark{b} \nl
13451+1232$^*$&   0.122 (1)&12.28&   S2 (1) \nl
13454$-$2956&   0.129 (1)&12.21&   S2 (1) \nl
14060+2919$^*$&   0.117 (1)&12.03&   H (1) \nl
14070+0525&   0.265 (1)&12.76&   S2 (1) \nl
 & & & \nl
14202+2615$^*$&   0.159 (1)&12.39&   H (1) \nl
14348$-$1447$^*$&   0.083 (1)&12.28&   L (1,5) \nl
14394+5332$^*$&   0.105 (1)&12.04&   S2 (1) \nl
15001+1433$^*$&   0.162 (1)&12.38&   S2 (1) \nl
15130$-$1958$^*$&   0.109 (1)&12.09&   S2 (1) \nl
15206+3342$^*$&   0.125 (1)&12.18&   H (1) \nl
15327+2340$^*$&   0.018 (1)&12.17&   L (1,5)\tablenotemark{c} \nl 
15462$-$0450&   0.100 (1)&12.16&   S1 (1) \nl
16090$-$0139$^*$&   0.134 (1)&12.49&   L (1) \nl
16300+1558&   0.242 (1)&12.63&   L (1) \nl
 & & & \nl
16333+4630$^*$&   0.191 (1)&12.35&   L (1) \nl
16468+5200$^*$&   0.150 (1)&12.02&   L (1) \nl
16474+3430$^*$&   0.111 (1)&12.11&   H (1) \nl
16487+5447$^*$&   0.104 (1)&12.12&   L (1) \nl
17028+5817$^*$&   0.106 (1)&12.10&   L (1) \nl
17044+6720$^*$&   0.135 (1)&12.13&   L (1) \nl
17068+4027$^*$&   0.179 (1)&12.30&   H (1) \nl
20414$-$1651$^*$&   0.086 (1)&12.14&   H (1) \nl
21329$-$2346$^*$&   0.125 (1)&12.09&   L (1) \nl
\cutinhead{New Data for Extended 1 Jy Sample (6 objects)}
14378$-$3651$^*$&   0.068 (1)&12.07&   L (1) \nl
17208$-$0014$^*$&   0.043 (1)&12.35&   H (1) \nl
17574+0629$^*$&   0.109 (1)&12.11&   H (1) \nl
18470+3233$^*$&   0.079 (1)&12.02&   H (1) \nl
19297$-$0406$^*$&   0.084 (1)&12.31&   H (1) \nl
20087$-$0308$^*$&   0.105 (1)&12.36&   L (1) \nl
\tablebreak
\cutinhead{Previously Published Data for 1 Jy Sample}
01572+0009&   0.163 (4)&12.53&   Q,S1 (6) \nl
05189$-$2524$^\dagger$&   0.042 (5)&12.07&   S2 (5) \nl
08559+1053&   0.149 (3)&12.16&   S2 (3) \nl
12265+0219&   0.159 (6)&12.73&   Q,S1 (6) \nl
13305$-$1739&   0.148 (2)&12.21&   S2 (2) \nl
21219$-$1757&   0.113 (3)&12.06&   S1 (3) \nl
22491$-$1808$^\dagger$&   0.076 (5)&12.09&   H (5) \nl
\tablenotetext{}{Column (1) -- Target name. FSC is the prefix given to the name 
of all sources in the {\it IRAS} Faint Source Catalog 
(c.f. Moshir et al. 1992). 
Galaxies used for spatial information (\S 5) 
are flagged with asterisk ($^*$)
and galaxies used only for spectral classification (\S 4.1)
are flagged with dagger ($^\dagger$).}
\tablenotetext{}{Column (2) -- Optical redshift.  Numbers in parentheses are references.}
\tablenotetext{}{Column (3) -- Infrared luminosity (see text).}
\tablenotetext{}{Column (4) -- Spectral type: H II = HII galaxy, L = LINER, 
S2 = Seyfert 2, S1 = Seyfert 1, Q = quasar.
Numbers in parentheses are references.}
\tablenotetext{a}
{Data for the 1~Jy and Extended 1~Jy objects are merged into a single list in  
subsequent tables.}
\tablenotetext{b}{This object was classified as a LINER by Veilleux et al. (1995).
        The reclassification of this object is due to a slight change in
        the [O~III]~$\lambda$5007/H$\beta$ ratio which may be attributable
        to differences in the extraction apertures between the two sets of
        data. See Section 4.1.}
\tablenotetext{c}{This object was classified as a Seyfert 2 by Veilleux et al. (1995).
        The present classification is considered more reliable because the
        data are of better quality and are less affected by centering errors.
        See Section 4.1.}
\tablerefs{For redshifts:\ 
(1) This work (see Paper{\ts}I for a comparison of our redshifts with previous 
values from the literature), (2) Allen et al. 1991, 
(3) Hill et al. 1988, 
(4) MacKenty \& Stockton 1984, 
(5) Sanders et al. 1988, (6) Schmidt 1963}
\tablerefs{For spectral type:\ 
(1) This work, (2) Allen et al. 1991, (3) Hill et al. 1988, (4) Low et al. 1988,
(5) Veilleux et al. 1995, (6) V\'eron--Cetty \& V\'eron 1989}
\enddata
\end{deluxetable}

\clearpage

\begin{deluxetable}{lcccccccccc}
\tablenum{3}
\tablewidth{0pt}
\tablecaption{Emission-Line Fluxes, Luminosities, and Electron Densities}
\tablehead{
\colhead{Name} &
\multicolumn{1}{c}{$F_{{\rm H}\alpha}$} &
\multicolumn{1}{c}{$EW_{{\rm H}\alpha}$} &
\multicolumn{1}{c}{$L_{{\rm H}\alpha}$} &
\multicolumn{1}{c}{H$\beta$} &
\multicolumn{1}{c}{[O III]} &
\multicolumn{1}{c}{[O I]} &
\multicolumn{1}{c}{[N II]} &
\multicolumn{1}{c}{[S II]} &
\multicolumn{1}{c}{[S II]} &
\colhead{$n_e$} \nl
\colhead{IRAS FSC} &
\multicolumn{1}{c}{} &
\multicolumn{1}{c}{} &
\multicolumn{1}{c}{log} &
\multicolumn{1}{c}{4861} &
\multicolumn{1}{c}{5007} &
\multicolumn{1}{c}{6300} &
\multicolumn{1}{c}{6583} &
\multicolumn{1}{c}{6716} &
\multicolumn{1}{c}{6731} &
\colhead{cm$^{-3}$} \nl 
\colhead{(1)} &
\multicolumn{1}{c}{(2)} &
\multicolumn{1}{c}{(3)} &
\multicolumn{1}{c}{(4)} &
\multicolumn{1}{c}{(5)} &
\multicolumn{1}{c}{(6)} &
\multicolumn{1}{c}{(7)} &
\multicolumn{1}{c}{(8)} &
\multicolumn{1}{c}{(9)} &
\multicolumn{1}{c}{(10)} &
\colhead{(11)}
}
\startdata
07598+6508 & 37.:: & 162:: & \nodata & \nodata & \nodata & \nodata & \nodata & \nodata & \nodata & \nodata \nl
08201+2801&0.36& 83& 8.30&0.21&0.14&0.05&0.47&0.19&0.10& \nodata\nl
08572+3915:NW&0.89& 72& 9.13&0.05&0.13&0.05&0.43&0.26&0.23& 350\nl
08572+3915:SE&0.40& 41& 8.70&&0.05&0.13&0.02&0.38&0.15& \nodata\nl
10091+4704&0.18& 29& 9.19&0.08&0.09&0.09&0.80&0.70s&\nodata & \nodata\nl
10494+4424&0.31& 37& 8.18&0.11&0.15&0.13&0.60&0.23&0.22& 550\nl
11029+3130&0.03&  9& 6.96&0.29&0.12&0.18&1.15&0.68s&\nodata & \nodata\nl
11119+3257&117.::&222::&\nodata &\nodata &\nodata &\nodata &\nodata &\nodata &\nodata &\nodata \nl
11223$-$1244&0.20& 40&10.33&0.02:&0.32&0.08&1.05&0.37&0.19& \nodata\nl
11506+1331&0.47& 89& 8.76&0.11&0.13&0.10&0.43&0.19&0.16& 230\nl
 & & & & & & & & & & \nl
11582+3020&0.14& 51& 8.59&0.15&0.17&0.13&0.73&0.20&0.22& 820\nl
12072$-$0444&1.70&175& 9.90&0.06&0.45&0.09&0.71&0.32&0.20& \nodata\nl
12112+0305&0.80& 65& 8.27&0.13&0.27&0.12&0.64&0.54s&\nodata & \nodata\nl
12540+5708&135.::&168::&\nodata &\nodata &\nodata &\nodata &\nodata &\nodata &\nodata &\nodata \nl
13218+0552&1.98:&391&\nodata &0.14:&0.29:&0.05:&0.25:&0.06:&0.04:& \nodata\nl
13342+3932&4.12&301&\nodata &0.07&0.28&0.02&0.53&0.03&0.03& 850\nl
13428+5608&8.21& 96& 8.97&0.10&0.36&0.13&1.02&0.31&0.27& 280\nl
13451+1232&2.80&248& 9.62&0.09&0.83&0.23&0.75&0.22&0.21& 450\nl
13454$-$2956&0.12& 26& 8.58&0.07&0.40&0.06&1.04&0.44s&\nodata & \nodata\nl
14060+2919&3.03&194& 9.25&0.14&0.11&0.04&0.46&0.17&0.16& 370\nl
 & & & & & & & & & & \nl
14070+0525&0.11& 26& 9.76&0.04&0.19&0.13&0.82&0.32&0.41& 410\nl
14202+2615&1.20&170& 9.03&0.16&0.13&0.04&0.46&0.17&0.13& 170\nl
14348$-$1447:SW&1.16& 91& 8.73&0.11&0.15&0.11&0.65&0.26&0.20& 140\nl
14348$-$1447:NE&0.49& 46& 8.40&0.10:&0.15&0.15&0.67&0.34&0.24&  30\nl
14378$-$3651&0.59& 39& 8.41&0.09&0.17&0.12&0.85&0.37s& \nodata & \nodata\nl
14394+5332&4.49&159& 9.36&0.13&0.43&0.20&0.94&0.43&0.19& \nodata\nl
15001+1433&0.64& 70& 8.89&0.13&0.48&0.06&0.65&0.22&0.23& 640\nl
15130$-$1958&1.05:& 98&10.05&0.04:&0.88:&0.08:&0.84:&0.28:s&\nodata & \nodata\nl
15206+3342&7.67&397& 9.58&0.17&0.52&0.03&0.22&0.08&0.06& 210\nl
15327+2340&1.09& 16& 7.29&0.11:&0.13&0.18&2.39&1.06s&\nodata & \nodata\nl
 & & & & & & & & & & \nl
15462$-$0450&5.6:&168&\nodata &0.18:&0.18:&0.01:&0.38:&0.07:s&\nodata & \nodata\nl
16090$-$0139&1.31&128& 9.66&0.07&0.08&0.08&0.77&0.24&0.16& \nodata\nl
16300+1558&0.22& 40& 8.89&0.12&0.15&0.11&0.60&0.32&0.19& \nodata\nl
16333+4630&0.49&100& 8.98&0.12&0.20&0.06&0.58&0.24&0.17&  10\nl
16468+5200:W&0.15& 47& 8.28&0.12&0.08&0.02&0.65&0.32&0.35& 820\nl
16468+5200:E&0.09& 30& 8.40&0.09&0.10&0.06:&0.69&0.43&0.38& 370\nl
16474+3430&1.28& 87& 8.33&0.24&0.26&0.06&0.48&0.16&0.15& 510\nl
16487+5447&2.92&185& 8.71&0.20&0.32&0.07&0.58&0.21&0.15&  50\nl
17028+5817:W&0.44& 72& 8.90&0.07&0.11&0.11&0.57&0.20&0.22& 740\nl
17028+5817:E&0.79&135& 8.39&0.18&0.20&0.02&0.28&0.18&0.13&  40\nl
 & & & & & & & & & & \nl \tablebreak
17044+6720&0.57& 99& 8.63&0.14&0.37&0.10&0.62&0.23&0.13& \nodata\nl
17068+4027&0.91&227& 9.71&0.08&0.17&0.03&0.34&0.22&0.15& \nodata\nl
17208$-$0014&1.79& 51& 8.96&0.06&0.05&0.05&0.46&0.19&0.14& 100\nl
17574+0629&2.16&278& 9.30&0.11&0.21&0.04&0.34&0.13&0.07& \nodata\nl
18470+3233&2.23&162& 8.99&0.11&0.13&0.05&0.44&0.17&0.11& \nodata\nl
19297$-$0406&0.64& 92& 8.86&0.08&0.05&0.08&0.56&0.20&0.16& 240\nl
20087$-$0308&0.45& 37& 8.85&0.05&0.10&0.18&1.02&0.52s&\nodata & \nodata\nl
20414$-$1651&0.37& 48& 8.37&0.11&0.08&0.11&0.54&0.18&0.14& 180\nl
21329$-$2346&0.50& 79& 8.30&0.17&0.17&0.10&0.61&0.45s&\nodata & \nodata\nl
\tablenotetext{}{Column (1) -- Target name.}
\tablenotetext{}{Column (2) -- Absolute flux of H$\alpha$ in units of 10$^{-14}$ erg
s$^{-1}$ cm$^{-2}$.} 
\tablenotetext{}{Column (3) -- Equivalent widths of the H$\alpha$ emission line in \AA derived in the rest frames of the objects.}
\tablenotetext{}{Column (4) -- Logarithms of the dereddened H$\alpha$ 
luminosities in units of
the bolometric solar luminosity (3.83 $\times$ 10$^{33}$ erg
s$^{-1}$).} 
\tablenotetext{}{Column (5) -- Flux of H$\beta$ relative to H$\alpha$. This
is the weighted average of the values obtained from the standard
plotting package of IRAF (`splot') and the fitting procedures,
with weights estimated from the relative uncertainties in the two
types of measurements (see \S 3).} 
\tablenotetext{}{Column (6) -- Flux of [O III] $\lambda$5007 relative to H$\alpha$.} 
\tablenotetext{}{Column (7) -- Flux of [O I] $\lambda$6300 relative to H$\alpha$.} 
\tablenotetext{}{Column (8) -- Flux of [N II] $\lambda$6583 relative to H$\alpha$.} 
\tablenotetext{}{Column (9) -- Flux of [S II] $\lambda$6716 relative to H$\alpha$. 
Entry followed by the letter ``s'' represents the sum of
[S II] $\lambda$6716 and [S II] $\lambda$6731 relative to H$\alpha$.} 
\tablenotetext{}{Column (10) -- Flux of [S II] $\lambda$6731 relative to H$\alpha$.} 
\tablenotetext{}{Column (11) -- Electron densities determined from the values of 
[S II] $\lambda$6716/[S II] $\lambda$6731 and the 
five-level-atom calculations of De~Robertis, Dufour, \& Hunt (1987).}
\tablecomments{
The emission line fluxes listed in Table 3 are normalized to
the observed flux of narrow H$\alpha$. In Seyfert 1s (see Table 5),
the H$\alpha$ flux, equivalent width, and luminosity refer to the sum
of the narrow and broad components of H$\alpha$.
The uncertainty on these measurements is typically of 5 -- 10\%, and
colons (:) indicate values with relative uncertainties of 20 -- 30\%. 
Double colons (::) indicate values with relative uncertainties larger than 30\% and
these objects were not considered in the analysis because of the large 
uncertainties in the data.} 
\enddata
\end{deluxetable}

\clearpage

\begin{deluxetable}{lccccccc}
\tablenum{4}
\tablewidth{0pt}
\tablecaption{Absorption Properties, Line Widths, and Continuum Levels}
\tablehead{
\colhead{Name} &
\multicolumn{3}{c}{{\it EW}(\AA)} &
\multicolumn{1}{c}{{\it FWHM}} &
\multicolumn{1}{c}{C4861} &
\colhead{{\it r}(C)} \nl
\cline{2-4}
\colhead{IRAS FSC} &
\multicolumn{1}{c}{H$\beta$} &
\multicolumn{1}{c}{Mg Ib} &
\multicolumn{1}{c}{Na ID} &
\multicolumn{1}{c}{[O III]} &
\multicolumn{1}{c}{} &
\multicolumn{1}{c}{} &
\colhead{} \nl 
\colhead{(1)} &
\multicolumn{1}{c}{(2)} &
\multicolumn{1}{c}{(3)} &
\multicolumn{1}{c}{(4)} &
\multicolumn{1}{c}{(5)} &
\multicolumn{1}{c}{(6)} &
\multicolumn{1}{c}{(7)} &
\colhead{(8)} 
}
\startdata
07598+6508&\nodata &\nodata &\nodata &\nodata &32.2&22.6&\nodata  \nl
08201+2801&\nodata &1.3 &4.6 & 510 &0.50&0.44&0.52 \nl
08572+3915:NW&\nodata  &0.7 &3.4 &  70 &1.07&1.23&0.17 \nl
08572+3915:SE&\nodata  &0.2 &2.7 & \nodata&0.77&0.99&0.20 \nl
10091+4704&0.8 &2.2 &3.2 & 360 &0.57&0.62&0.28 \nl
10494+4424&8.2 &0.1 &4.1 & \nodata&0.52&0.83&0.56 \nl
11029+3130&5.6:&1.5 &2.7 & \nodata&0.35&0.34&0.89 \nl
11119+3257&\nodata &\nodata &\nodata &\nodata &2.03&4.74&\nodata  \nl
11223$-$1244&2.8 &2.3 &6.9 & 960 &0.36&0.50&0.10 \nl
11506+1331&1.1 &0.8 &4.9 &  80 &0.44&0.53&0.39 \nl
& & & \nl
11582+3020&0.5:&0.2 &4.0 & 350 &0.33&0.27&0.38 \nl
12072$-$0444&0.5:&0.3 &2.1 & 500 &0.81&0.97&0.22 \nl
12112+0305&1.2:&1.5 &4.0 & \nodata&1.26&1.23&0.39 \nl
12540+5708&\nodata&\nodata &\nodata &\nodata &85.0&75.1&\nodata  \nl
13218+0552&\nodata  &2.4 &0.8 &1430 &0.59&0.51&\nodata  \nl
13342+3932&0.5:&0.8 &2.4 & 420 &1.87&1.37&\nodata  \nl
13428+5608&0.3 &0.7 &4.9 & 480 &7.76&8.55&0.33 \nl
13451+1232&0.2 &1.5 &0.5 &1250 &0.80&1.13&0.40 \nl
13454$-$2956&\nodata  &0.2 &1.9 &  70 &0.47&0.46&0.21 \nl
14060+2919&0.7 &1.2 &2.8 & 200 &1.56&1.56&0.41 \nl
& & & \nl
14070+0525&0.7 &1.5 &4.6 & \nodata&0.34&0.42&0.15 \nl
14202+2615&0.3 &1.4 &3.5 & 160 &0.77&0.71&0.42 \nl
14348$-$1447:SW&3.4 &0.6 &4.8 & 440 &1.00&1.28&0.42 \nl
14348$-$1447:NE&3.7 &0.4 &3.9 & 100 &0.95&1.08&0.36 \nl
14378$-$3615&0.2 &3.3 &7.8 & \nodata&1.20&1.52&0.36 \nl
14394+5332&0.4 &0.2 &3.9 &1870 &2.41&2.83&0.46 \nl
15001+1433&0.6 &1.2 &4.2 & 280 &0.96&0.91&0.38 \nl
15130$-$1958&0.2:&0.2 &3.6 &1100 &0.92&1.07&0.13 \nl
15206+3342&0.1 &0.2 &1.9 &  60 &2.41&1.93&0.38 \nl
15327+2340&2.6:&2.7 &5.8 & \nodata&4.50&6.93&0.54 \nl
& & & \nl
15462$-$0450&\nodata  &0.3 &0.2:&1560 &3.24&3.33&\nodata  \nl
16090$-$0139&0.5 &1.8 &4.4 & 150 &0.79&1.02&0.28 \nl
16300+1558&2.1 &1.0 &5.5 & 120 &0.56&0.54&0.35 \nl
16333+4630&0.7 &0.4 &3.4 & 450 &0.45&0.49&0.42 \nl
16468+5200:W&\nodata  &\nodata  &4.3 & \nodata&0.33&0.32&0.37 \nl
16468+5200:E&\nodata  &\nodata  &\nodata  & \nodata&0.28&0.31&0.30 \nl
16474+3430&2.2 &0.4 &1.6 & \nodata&1.92&1.47&0.52 \nl
16487+5447&0.1 &1.4 &1.2 & 340 &2.16&1.58&0.45 \nl
17028+5817:W&0.9 &1.5 &5.9 &  80 &0.46&0.62&0.31 \nl
17028+5817:E&0.9 &2.2 &2.5 & 120 &0.61&0.59&0.48 \nl
\tablebreak
17044+6720&0.5 &0.9 &0.8 & 280 &0.61&0.58&0.40 \nl
17068+4027&0.9 &1.7 &3.9 & 440 &0.35&0.40&0.26 \nl
17208$-$0014&2.1 &1.1 &8.4 &  60 &1.77&3.49&0.34 \nl
17574+0629&0.8 &1.5 &0.7 & 240 &0.70&0.78&0.35 \nl
18470+3233&0.1 &0.3 &2.9 & 230 &1.27&1.38&0.35 \nl
19297$-$0406&0.8 &1.2 &4.5 & 190 &0.42&0.70&0.38 \nl
20087$-$0308&2.2:&4.7 &9.8 & 160 &0.73&1.19&0.39 \nl
20414$-$1651&1.1 &0.5 &3.6 & \nodata &0.57&0.77&0.40 \nl
21329$-$2346&3.1 &2.2 &2.9 & 290 &0.87&0.63&0.37 \nl
\tablenotetext{}{Column (1) -- Target name.}
\tablenotetext{}{Column (2) -- Equivalent widths of the H$\beta$ 
absorption feature in \AA~derived from the fitting method and in the
rest frames of the objects.
The typical uncertainties on these measurements are about 10 -- 20\%,
and for some of the objects with strong H$\beta$ emission line 
they are $\sim$30\% (values flagged with a colon).}
\tablenotetext{}{Column (3) -- Equivalent widths of the Mg Ib $\lambda$5176 absorption
feature in \AA derived in the rest frames of the objects. 
The typical uncertainties on these measurements are about 10 -- 20\%,
and for some of the objects they are $\sim$30\% (values flagged with a colon).}
\tablenotetext{}{Column (4) -- Equivalent widths of the H$\beta$ absorption feature in
\AA derived in the rest frames of the objects. The typical uncertainties on these measurements are about 10 -- 20\%,
and for some of the objects they are $\sim$30\% (values flagged with a colon).}
\tablenotetext{}{Column (5) -- Line widths of [O III] $\lambda$5007 in km s$^{-1}$
after correction for the finite instrument resolution of the data
using the quadrature method (see \S 4.4).}
\tablenotetext{}{Column (6) -- Observed continuum levels near 4861 \AA~in units of 10$^{-16}$ erg
s$^{-1}$ cm$^{-2}$ \AA$^{-1}$.}
\tablenotetext{}{Column (7) -- Observed continuum levels near 6563 \AA~in units of 10$^{-16}$ erg
s$^{-1}$ cm$^{-2}$ \AA$^{-1}$.}
\tablenotetext{}{Column (8) -- Dereddened continuum ratios ([C6563/C4861]$_0$).}
\enddata
\end{deluxetable}

\clearpage

\begin{deluxetable}{lcccccccccc}
\tablenum{5}
\tablewidth{0pt}
\tablecaption{Observed and Dereddened Line Ratios and Spectral Classification}
\tablehead{
\colhead{Name} &
\multicolumn{1}{c}{${\rm H\alpha \over H\beta}$} &
\multicolumn{1}{c}{$E(B-V)$} &
\multicolumn{1}{c}{${\rm [O III] \over H\beta}$} &
\multicolumn{1}{c}{${\rm [N II] \over H\alpha}$} &
\multicolumn{1}{c}{${\rm [S II] \over H\alpha}$} &
\multicolumn{1}{c}{${\rm [O I] \over H\alpha}$} &
\multicolumn{4}{c}{Spectral Type} \nl
\cline{8-11}
\colhead{IRAS FSC} &
\multicolumn{1}{c}{log} &
\multicolumn{1}{c}{} &
\multicolumn{1}{c}{log} &
\multicolumn{1}{c}{log} &
\multicolumn{1}{c}{log} &
\multicolumn{1}{c}{log} &
\multicolumn{1}{c}{[N II]} &
\multicolumn{1}{c}{[S II]} &
\multicolumn{1}{c}{[O I]} &
\colhead{Adopt} \nl 
\colhead{(1)} &
\multicolumn{1}{c}{(2)} &
\multicolumn{1}{c}{(3)} &
\multicolumn{1}{c}{(4)} &
\multicolumn{1}{c}{(5)} &
\multicolumn{1}{c}{(6)} &
\multicolumn{1}{c}{(7)} &
\multicolumn{1}{c}{(8)} &
\multicolumn{1}{c}{(9)} &
\multicolumn{1}{c}{(10)} &
\colhead{(11)}
}
\startdata
07598+6508&\nodata &\nodata &\nodata  &\nodata  &\nodata  &\nodata  &   &   &   & S1 \nl
       &\nodata &    &\nodata  &\nodata  &\nodata  &\nodata  &   &   &   &    \nl
08201+2801&0.68&0.53&--0.19&--0.32&--0.54&--1.32&H&H&H&H \nl
       &0.45&   &--0.21&--0.33&--0.55&--1.29&   &   &   &    \nl
08572+3915:NW&1.33&1.94&$\ $0.46&--0.37&--0.31&--1.30&H&  L&  L&  L: \nl
       &0.49&   &$\ $0.36&--0.38&--0.37&--1.20&   &   &   &    \nl
08572+3915:SE&1.29&1.85&$\ $0.42&$\ $0.00&--0.28&--1.78&  L&  L&H&  L: \nl
       &0.49&   &$\ $0.33&--0.01&--0.34&--1.69&   &   &   &    \nl
10091+4704&1.08&1.36&$\ $0.05&--0.10&--0.16&--1.06&  L&H&  L&  L: \nl
       &0.49&   &--0.02&--0.10&--0.20&--0.99&   &   &   &    \nl
10494+4424&0.94&1.04&$\ $0.12&--0.22&--0.35&--0.90&  L&  L&  L&  L \nl
       &0.49&   &$\ $0.07&--0.23&--0.38&--0.85&   &   &   &    \nl
11029+3130&0.53&0.10&--0.37&$\ $0.06&--0.17&--0.74&  L&  L&  L&  L \nl
       &0.49&   &--0.38&$\ $0.06&--0.17&--0.74&   &   &   &    \nl
11119+3257&\nodata &\nodata &--0.04&\nodata  &\nodata  &\nodata  &   &   &   & S1 \nl
       &0.49&   &\nodata  &\nodata  &\nodata  &\nodata  &   &   &   &    \nl
11223$-$1244&1.65&2.68& 1.16&$\ $0.02&--0.25&--1.10& S2& S2& S2& S2 \nl
       &0.49&   & 1.03&$\ $0.01&--0.34&--0.96&   &   &   &    \nl
11506+1331&0.98&1.13&$\ $0.08&--0.37&--0.46&--0.98&H&H&  L&H: \nl
       &0.45&   &$\ $0.02&--0.38&--0.50&--0.92&   &   &   &    \nl
11582+3020&0.82&0.76&$\ $0.04&--0.13&--0.38&--0.89&  L&  L&  L&  L \nl
       &0.49&   &$\ $0.00&--0.14&--0.40&--0.85&   &   &   &    \nl
12072$-$0444&1.23&1.71&$\ $0.88&--0.15&--0.29&--1.07& S2& S2& S2& S2 \nl
       &0.49&   &$\ $0.80&--0.15&--0.34&--0.98&   &   &   &    \nl
12112+0305&0.89&0.92&$\ $0.32&--0.19&--0.27&--0.93&  L&  L&  L&  L \nl
       &0.49&   &$\ $0.28&--0.20&--0.29&--0.89&   &   &   &    \nl
12540+5708&\nodata &\nodata &\nodata  &\nodata  &\nodata  &\nodata  &   &   &   & S1 \nl
       &0.49&   &\nodata  &\nodata  &\nodata  &\nodata  &   &   &   &    \nl
13218+0552&0.85&0.83&$\ $0.31&--0.61&--0.99&--1.35&&&& S1 \nl
       &0.49&   &$\ $0.27&--0.61&--1.01&--1.30&   &   &   &    \nl
13342+3932&1.17&1.57&$\ $0.62&--0.28&--1.22&--1.77& &&& S1 \nl
       &0.49&   &$\ $0.55&--0.29&--1.27&--1.69&   &   &   &    \nl
13428+5608&1.02&1.22&$\ $0.57&$\ $0.01&--0.24&--0.88& S2& S2& S2& S2 \nl
       &0.49&   &$\ $0.51&$\ $0.00&--0.27&--0.82&   &   &   &    \nl
13451+1232&1.04&1.26&$\ $0.96&--0.12&--0.37&--0.65& S2& S2& S2& S2 \nl
       &0.49&   &$\ $0.90&--0.13&--0.41&--0.58&   &   &   &    \nl
13454$-$2956&1.16&1.54&$\ $0.76&$\ $0.02&--0.36&--1.23& S2& S2& S2& S2 \nl
       &0.49&   &$\ $0.68&$\ $0.01&--0.41&--1.15&   &   &   &    \nl
14060+2919&0.84&0.90&--0.12&--0.34&--0.49&--1.42&H&H&H&H \nl
       &0.45&   &--0.17&--0.34&--0.51&--1.38&   &   &   &    \nl
14070+0525&1.39&2.08&$\ $0.67&--0.09&--0.14&--0.90& S2& S2& S2& S2 \nl
       &0.49&   &$\ $0.57&--0.10&--0.20&--0.79&   &   &   &    \nl
14202+2615&0.80&0.79&--0.10&--0.34&--0.53&--1.39&H&H&H&H \nl
       &0.45&   &--0.14&--0.34&--0.55&--1.35&   &   &   &    \nl
14348$-$1447:SW&0.99&1.15&$\ $0.17&--0.18&--0.23&--0.83&  L&  L&  L&  L \nl
       &0.49&   &$\ $0.11&--0.18&--0.27&--0.77&   &   &   &    \nl
14348$-$1447:NE&0.97&1.11&$\ $0.15&--0.18&--0.34&--0.98&  L&  L&  L&  L \nl
       &0.49&   &$\ $0.09&--0.19&--0.38&--0.92&   &   &   &    \nl
14378$-$3651&1.04&1.26&$\ $0.28&--0.07&--0.44&--0.94&  L&H&  L&  L: \nl
       &0.49&   &$\ $0.21&--0.07&--0.48&--0.87&   &   &   &    \nl
14394+5332&0.89&0.93&$\ $0.53&--0.03&--0.21&--0.71& S2& S2& S2& S2 \nl
       &0.49&   &$\ $0.49&--0.03&--0.24&--0.66&   &   &   &    \nl
15001+1433&0.89&0.91&$\ $0.57&--0.19&--0.35&--1.21& S2& S2& S2& S2 \nl
       &0.49&   &$\ $0.53&--0.19&--0.37&--1.16&   &   &   &    \nl
15130$-$1958&1.45&2.23& 1.40&--0.08&--0.55&--1.13& S2& S2& S2& S2 \nl
       &0.49&   & 1.29&--0.09&--0.62&--1.01&   &   &   &    \nl
15206+3342&0.78&0.75&$\ $0.49&--0.65&--0.86&--1.60&H&H&H&H \nl
       &0.45&   &$\ $0.45&--0.65&--0.88&--1.56&   &   &   &    \nl
15327+2340&0.95&1.05&$\ $0.06&$\ $0.38&$\ $0.02&--0.74&  L&  L&  L&  L \nl
       &0.49&   &$\ $0.01&$\ $0.38&--0.01&--0.68&   &   &   &    \nl
15462$-$0450&0.75&0.60&$\ $0.00&--0.42&--1.15&--1.84&&&& S1 \nl
       &0.49&   &--0.02&--0.42&--1.17&--1.81&   &   &   &    \nl
16090$-$0139&1.16&1.55&$\ $0.08&--0.11&--0.39&--1.11&  L&H&  L&  L: \nl
       &0.49&   &$\ $0.01&--0.12&--0.44&--1.03&   &   &   &    \nl
16300+1558&0.92&1.00&$\ $0.09&--0.22&--0.29&--0.97&  L&  L&  L&  L \nl
       &0.49&   &$\ $0.04&--0.23&--0.32&--0.92&   &   &   &    \nl
16333+4630&0.91&0.96&$\ $0.20&--0.24&--0.38&--1.19&  L&  L&  L&  L \nl
       &0.49&   &$\ $0.16&--0.24&--0.41&--1.15&   &   &   &    \nl
16468+5200:W&0.92&0.99&--0.19&--0.19&--0.18&--1.69&  L&  L&H&  L: \nl
       &0.49&   &--0.24&--0.20&--0.21&--1.64&   &   &   &    \nl
16468+5200:E&1.07&1.33&$\ $0.06&--0.16&--0.09&--1.24&  L&  L&  L&  L \nl
       &0.49&   &--0.01&--0.17&--0.13&--1.18&   &   &   &    \nl
16474+3430&0.62&0.39&$\ $0.04&--0.32&--0.51&--1.25&H&H&  L&H: \nl
       &0.45&   &$\ $0.03&--0.32&--0.52&--1.23&   &   &   &    \nl
16487+5447&0.70&0.48&$\ $0.20&--0.24&--0.44&--1.16&  L&H&  L&  L: \nl
       &0.49&   &$\ $0.18&--0.24&--0.46&--1.14&   &   &   &    \nl
17028+5817:W&1.13&1.47&$\ $0.16&--0.25&--0.38&--0.97&  L&H&  L&  L: \nl
       &0.49&   &$\ $0.09&--0.25&--0.42&--0.90&   &   &   &    \nl
17028+5817:E&0.76&0.70&$\ $0.05&--0.56&--0.51&--1.64&H&H&H&H \nl
       &0.45&   &$\ $0.01&--0.56&--0.53&--1.61&   &   &   &    \nl
17044+6720&0.87&0.86&$\ $0.43&--0.21&--0.44&--0.99&  L&  L&  L&  L \nl
       &0.49&   &$\ $0.39&--0.21&--0.46&--0.95&   &   &   &    \nl
17068+4027&1.10&1.48&$\ $0.33&--0.46&--0.44&--1.47&H&H&H&H \nl
       &0.45&   &$\ $0.26&--0.47&--0.48&--1.40&   &   &   &    \nl
17207$-$0014&1.21&1.75&--0.08&--0.34&--0.48&--1.31&H&H&H&H \nl
       &0.45&   &--0.16&--0.35&--0.54&--1.22&   &   &   &    \nl
17574+0629&0.95&1.15&$\ $0.27&--0.48&--0.69&--1.44&H&H&H&H \nl
       &0.45&   &$\ $0.21&--0.48&--0.73&--1.38&   &   &   &    \nl
18470+3233&0.95&1.13&$\ $0.06&--0.36&--0.56&--1.34&H&H&H&H \nl
       &0.45&   &$\ $0.01&--0.37&--0.59&--1.29&   &   &   &    \nl
19297$-$0406&1.10&1.49&--0.16&--0.25&--0.44&--1.09&H&H&  L&H: \nl
       &0.45&   &--0.23&--0.26&--0.49&--1.01&   &   &   &    \nl
20087$-$0308&1.11&1.43&$\ $0.26&$\ $0.01&--0.28&--0.75&  L&  L&  L&  L \nl
       &0.49&   &$\ $0.17&$\ $0.00&--0.34&--0.65&   &   &   &    \nl
20414$-$1651&0.98&1.21&--0.11&--0.27&--0.50&--0.97&H&H&  L&H: \nl
       &0.45&   &--0.17&--0.27&--0.54&--0.90&   &   &   &    \nl
21329$-$2346&0.78&0.67&$\ $0.00&--0.21&--0.35&--0.98&  L&  L&  L&  L \nl
       &0.49&   &--0.03&--0.22&--0.37&--0.95&   &   &   &    
\tablenotetext{}{Column (1) -- Target name.}
\tablenotetext{}{Column (2) -- Logarithm of H$\alpha$/H$\beta$.}
\tablenotetext{}{Column (3) -- Color excesses determined from the emission-line Balmer decrements (cf. \S 4.2).}
\tablenotetext{}{Column (4) -- Logarithm of reddening-corrected [O~III] $\lambda$5007/H$\beta$.}
\tablenotetext{}{Column (5) -- Logarithm of reddening-corrected [N~II] $\lambda$6583/H$\alpha$.}
\tablenotetext{}{Column (6) -- Logarithm of reddening-corrected [S~II] $\lambda\lambda$6716, 6731 /H$\alpha$.}
\tablenotetext{}{Column (7) -- Logarithm of reddening-corrected [O~I] $\lambda$6300/H$\alpha$.}
\tablenotetext{}{Columns (8) through (11) -- Optical spectral types determined from the
diagrams of $\ $ Veilleux \& Osterbrock (1987) involving [N~II] $\lambda$6583
[column (8)], [S~II] $\lambda\lambda$6716, 6731 [column (9)], and
[O~I] $\lambda$6300 [column (10)]. Column (11) lists the adopted
spectral type. Meaning of the symbols: H = H~II galaxies, L = LINER
(`low-ionization nuclear emission-line regions'), S2 = Seyfert 2 galaxies,
and S1 = Seyfert 1 galaxies. A colon (:) indicates galaxies with line ratios
that do not correspond to the same spectral type in all three diagrams of
Veilleux \& Osterbrock (1987). See \S 4.1 for further detail.}
\tablecomments{The emission line ratios listed in Table 3 are normalized to
the observed flux of narrow H$\alpha$ or H$\beta$. In Seyfert 1s,
the ratios are normalized to the sum
of the narrow and broad components of H$\alpha$ or H$\beta$. These ratios are only listed
for illustrative purposes since they cannot be used for spectral
classification.
The uncertainty on the emission line ratios is typically $\pm$10\%.
For entries with a colon (:) the uncertainty is approximately $\pm$25\%.}
\enddata
\end{deluxetable}

\clearpage

\begin{deluxetable}{lcccc}
\tablenum{6}
\tablewidth{0pt}
\tablecaption{Spectral Classification as a Function of {\it L}$_{\rm ir}$}
\tablehead{
\colhead{log $\left( L_{ir} \over L_\odot \right)$ } &
\multicolumn{1}{c}{$<$11\tablenotemark{a}} &
\multicolumn{1}{c}{11--11.99\tablenotemark{a}} &
\multicolumn{1}{c}{12--12.29} &
\multicolumn{1}{c}{12.3--12.9} \nl
}
\startdata
 & & \nl
Number&29&68&27&20 \nl\nl\nl
H II region (\%)&62&53&33&20 \nl\nl
AGN-like (\%)&38&47&67&80 \nl
{\hskip 0.3cm}Seyfert 1 (\%)&0&2&4&25 \nl
{\hskip 0.3cm}Seyfert 2 (\%)&7&13&22&20 \nl
{\hskip 0.3cm}LINER (\%)&31&32&41&35 \nl
\tablenotetext{a}{Data from luminous infrared galaxies in the BGS studied by VKSMS.}
\enddata
\end{deluxetable}

\clearpage

\begin{deluxetable}{lcccccccccccccccc}
\small
\tablenum{7}
\tablecaption{Radial Profiles of the Spectral Parameters}
\tablehead{
\colhead{$\ \ \ $} &
\multicolumn{1}{c}{Type\tablenotemark{a}} &
\multicolumn{3}{c}{0 -- 1 kpc} &
\multicolumn{1}{c}{} &
\multicolumn{3}{c}{1 -- 2 kpc} &
\multicolumn{1}{c}{} &
\multicolumn{3}{c}{2 -- 4 kpc} &
\multicolumn{1}{c}{} &
\multicolumn{3}{c}{4 -- 8 kpc} \nl
\cline{3-5} \cline{7-9} \cline{11-13} \cline{15-17}
\colhead{$\ \ \ $} &
\multicolumn{1}{c}{} &
\multicolumn{1}{c}{Mean} &
\multicolumn{1}{c}{Median} &
\multicolumn{1}{c}{No.\tablenotemark{b}} &
\multicolumn{1}{c}{} &
\multicolumn{1}{c}{Mean} &
\multicolumn{1}{c}{Median} &
\multicolumn{1}{c}{No.} &
\multicolumn{1}{c}{} &
\multicolumn{1}{c}{Mean} &
\multicolumn{1}{c}{Median} &
\multicolumn{1}{c}{No.} &
\multicolumn{1}{c}{} &
\multicolumn{1}{c}{Mean} &
\multicolumn{1}{c}{Median} &
\colhead{No.} \nl
}
\startdata
\sidehead{$E(B-V)$}
&H II&1.16$\pm$0.42&1.15&6&&1.30$\pm$0.56&1.27&6&&0.91$\pm$0.35&0.82&11&&0.80$\pm$0.48&0.97&5 \nl
&LINER&1.29$\pm$0.52&1.43&9&&1.41$\pm$0.67&1.27&8&&1.15$\pm$0.52&1.13&12&&0.74$\pm$0.57&0.63&4 \nl
&Sey 2&1.64$\pm$0.92&1.25&3&&1.30$\pm$0.48&1.10&3&&1.09$\pm$0.96&0.92&6&&0.98$\pm$0.98&0.98&1 \nl
\sidehead{$FWHM$([O III])}
&H II&130$\pm$110&190&6&&160$\pm$160&160&6&&320$\pm$230&310&11&&470$\pm$440&460&5 \nl
&LINER&100$\pm$130&40&9&&260$\pm$240&190&8&&230$\pm$220&190&13&&310$\pm$290&280&4 \nl
&Sey 2&1140$\pm$740&1130&3&&1200$\pm$620&1040&3&&900$\pm$630&650&6&&200$\pm$290&200&2 \nl
\sidehead{$EW$(H$\beta$)}
&H II&1.0$\pm$0.8&1.0&6&&1.7$\pm$1.6&1.5&6&&2.0$\pm$1.9&1.8&12&&1.8$\pm$1.2&1.9&3 \nl
&LINER&1.9$\pm$2.5&1.2&9&&2.3$\pm$2.1&1.9&8&&2.1$\pm$2.0&1.1&13&&2.7$\pm$2.1&2.5&4 \nl
&Sey 2&0.3$\pm$0.1&0.3&3&&0.4$\pm$0.3&0.2&3&&0.8$\pm$0.7&0.7&6&&2.5$\pm$2.6&2.5&2 \nl
\sidehead{$EW$(Mg Ib)}
&H II&1.2$\pm$1.0&0.7&6&&1.2$\pm$1.3&0.8&6&&0.9$\pm$0.7&0.6&12&&0.7$\pm$0.6&0.6&5 \nl
&LINER&2.1$\pm$1.8&1.9&9&&1.6$\pm$1.3&1.3&8&&1.6$\pm$1.0&1.6&13&&3.0$\pm$3.1&1.8&4 \nl
&Sey 2&0.4$\pm$0.3&0.4&3&&0.7$\pm$0.5&0.7&3&&1.2$\pm$0.8&0.9&6&&1.1$\pm$0.1&1.1&2 \nl
\sidehead{(C6563/C4861)$_0$}
&H II&0.40$\pm$0.05&0.39&6&&0.36$\pm$0.11&0.36&6&&0.44$\pm$0.09&0.45&11&&0.42$\pm$0.29&0.38&5 \nl
&LINER&0.39$\pm$0.22&0.39&9&&0.35$\pm$0.16&0.39&8&&0.45$\pm$0.34&0.36&12&&0.61$\pm$0.26&0.58&4 \nl
&Sey 2&0.30$\pm$0.19&0.33&3&&0.33$\pm$0.15&0.38&3&&0.50$\pm$0.39&0.41&6&&0.29$\pm$0.11&0.29&2 \nl
\sidehead{log $\left( L_{{\rm H} \alpha} / L_\odot \right)_0$}
&H II&8.65$\pm$0.28&8.69&6&&8.54$\pm$0.25&8.55&6&&8.77$\pm$0.57&8.35&11&&8.90$\pm$0.27&8.94&5 \nl
&LINER&8.65$\pm$0.42&8.54&9&&8.70$\pm$0.69&8.36&8&&8.70$\pm$0.62&8.44&12&&8.75$\pm$0.79&7.81&4 \nl
&Sey 2&9.89$\pm$0.64&9.20&3&&8.97$\pm$0.34&9.12&3&&9.38$\pm$0.89&8.37&6&&8.74$\pm$0.13&8.74&2 \nl
\sidehead{$EW$(H$\alpha$)}
&H II&125$\pm$81&99&6&&109$\pm$79&89&6&&127$\pm$115&77&12&&140$\pm$101&110&5 \nl
&LINER&78$\pm$53&65&9&&68$\pm$46&58&8&&62$\pm$37&50&13&&63$\pm$19&63&4 \nl
&Sey 2&129$\pm$37&112&3&&98$\pm$59&72&3&&61$\pm$21&71&6&&37$\pm$33&37&2 \nl
\tablenotetext{a}{Nuclear spectral type.}
\tablenotetext{b}{Number of objects used for statistics.}
\enddata
\end{deluxetable}